\documentclass[twocolumn]{autart}    

\usepackage{graphicx}          

\usepackage{cite}
\usepackage{amsmath,amssymb,amsfonts}
\usepackage{mathabx}
\usepackage{algorithm}
\usepackage{algpseudocode}
\usepackage{textcomp}
\usepackage[utf8]{inputenc}
\usepackage[T1]{fontenc}
\usepackage{balance}
\usepackage[rightcaption]{sidecap}
\usepackage{import}
\usepackage{textcomp}
\usepackage[dvipsnames]{xcolor}
\usepackage{slashed}
\usepackage{subfigure}
\usepackage{enumitem}
\usepackage{tabularx}
\usepackage{wrapfig}
\usepackage{bm}

\usepackage{easyReview}

\newcommand{\bbN}{\mathbb{N}}

\newcommand{\bbR}{\mathbb{R}}

\newcommand{\calA}{\mathcal{A}}
\newcommand{\calB}{\mathcal{B}}
\newcommand{\calC}{\mathcal{C}}

\newcommand{\calE}{\mathcal{E}}

\newcommand{\calG}{\mathcal{G}}

\newcommand{\calK}{\mathcal{K}}

\newcommand{\calN}{\mathcal{N}}

\newcommand{\calP}{\mathcal{P}}

\newcommand{\calR}{\mathcal{R}}

\newcommand{\calU}{\mathcal{U}}
\newcommand{\calV}{\mathcal{V}}
\newcommand{\calW}{\mathcal{W}}
\newcommand{\calX}{\mathcal{X}}

\definecolor{myblue}{rgb}{0, 0.4471, 0.7412}
\definecolor{mygreen}{rgb}{0.4667, 0.6745, 0.1882}
\definecolor{myred}{rgb}{0.8510, 0.3255, 0.0980}
\definecolor{myviolet}{rgb}{0.4941, 0.1843, 0.5569}

\begin{document}

\begin{frontmatter}

\title{Parallelized Robust Distributed Model Predictive Control in the Presence of Coupled State Constraints\thanksref{footnoteinfo1}} 

\thanks[footnoteinfo1]{This work was supported by the ERC Consolidator Grant LEAFHOUND, the Swedish Research Council (VR) and the Knut och Alice Wallenberg Foundation (KAW).}

\author{Adrian Wiltz}\ead{wiltz@kth.se},    
\author{Fei Chen}\ead{fchen@kth.se},               
\author{Dimos V. Dimarogonas}\ead{dimos@kth.se}  

\address{Division of Decision and Control Systems, KTH Royal Institute of Technology, Stockholm, Sweden}


\begin{abstract}                          
	In this paper, we present a robust distributed model predictive control (DMPC) scheme for dynamically decoupled nonlinear systems which are subject to state constraints, coupled state constraints and input constraints. In the proposed control scheme, all subsystems solve their local optimization problem in parallel and neighbor-to-neighbor communication suffices. The approach relies on consistency constraints which define a neighborhood around each subsystem's reference trajectory where the state of the subsystem is guaranteed to stay in. Contrary to related approaches, the reference trajectories are improved consecutively. In order to ensure the controller's robustness against bounded uncertainties, we employ tubes. The presented approach can be considered as a time-efficient alternative to the well-established sequential DMPC. In the end, we briefly comment on an iterative extension. The effectiveness of the proposed DMPC scheme is demonstrated with simulations, and its performance is compared to other DMPC schemes. 
\end{abstract}

\end{frontmatter}

\setlength{\parskip}{0.2cm}


\section{INTRODUCTION}

Model predictive control (MPC) algorithms are successfully employed in a wide range of applications and there exists a broad body of literature. The fact that hard constraints on states and inputs can be directly incorporated into the controller and a performance criterion can be taken into account by means of solving an optimization problem significantly contributes to their popularity. A basic approach for proving recursive feasibility and convergence of the optimization problems has been developed in~\cite{chen1998} for continuous-time nonlinear systems, and is the basis for numerous MPC algorithms proposed since then. Similar results are derived for discrete-time systems in~\cite{DeNicolao1998}, and an overview can be found in~\cite{gruene2011,mayne2000, mayne2014}. 

\subsection{Distributed MPC}
Since the presentation of initial distributed model predictive control (DMPC) schemes \cite{Camponogara2002}, their development became a thriving branch in the research on MPC. The motivation behind the development of DMPC is that centralized MPC \cite{mayne2000} becomes computationally intractable for large-scale systems, and a reliable communication with a central control-unit is difficult to realize in the case of spatially distributed systems \cite{Frejo2012,Hermans2012}.

In~\cite{mueller2017}, the methods by which distributed MPC algorithms compute control input trajectories are classified into four groups: iterative methods, sequential methods, methods employing consistency constraints, and approaches based on robustness considerations. In iterative methods, the local controllers exchange the solutions to their local optimization problems several times among each other until they converge. In sequential approaches, local optimization problems of neighboring subsystems are not evaluated in parallel but one after another. In algorithms based on consistency constraints, neighboring subsystems exchange reference trajectories and guarantee to stay in their neighborhood. Other DMPC algorithms consider the neighbors' control decisions as a disturbance. Examples can be found in \cite{mueller2017}. As remarked in \cite{Negenborn2014}, the task of distributing MPC algorithms is too complex in order to solve it with one single approach. Instead, for various types of centralized MPC problems, distributed controllers have been taylored. A broad collection of notable DMPC algorithms can be found in \cite{Negenborn2014a}. 

Especially the distribution of MPC problems subject to coupled state constraints turned out to be complicated \cite{Keviczky2006}, and most available DMPC schemes that are capable of handling them cannot avoid a sequential scheme \cite{Kuwata2011,Nikou2019,Richards2007,Trodden2010}. However, sequential schemes have the drawback that the computation of the control input for all subsystems becomes very time-consuming for highly connected networks. A notable exception that does not rely on a sequential scheme can be found in \cite{farina2012}, where a consistency constraint approach is used instead. This admits that even in the presence of coupled state constraints all subsystems can solve their local optimization problem in parallel and still retain recursive feasibility. In \cite{farina2014}, this approach is transferred to a continuous-time setup, and in \cite{Riverso2012,Riverso2013,Riverso2014a}, it is extended to a so-called plug-and-play MPC algorithm. However, \cite{farina2012,farina2014} employ fixed reference trajectories, i.e., reference trajectories that may not be modified once defined. This is limiting and restricts the possibility to optimize the system's performance significantly. In this paper, we overcome this limitation for dynamically decoupled systems subject to coupled state constraints.

\subsection{Contributions}
We develop a generic DMPC scheme based on consistency constraints that allows for the parallelized evaluation of the local optimization problems in the presence of coupled state constraints. In contrast to the consistency constraint based approaches \cite{farina2012,farina2014} with fixed reference trajectories, we show that in the case of dynamically decoupled systems this limitation can be overcome. In particular, we show that recursive feasibility and asymptotic stability can be still obtained when invoking a less restrictive assumption on reference trajectories and consistency constraints. It allows the reference trajectories and consistency constraints to be updated at every time-step. As nominal MPC subject to constraints is sensitive to model uncertainties and disturbances, we formulate the proposed DMPC scheme via a constraint-tightening approach using robust tubes. 

The proposed DMPC scheme can be considered as a time-efficient and scalable alternative to sequential DMPC \cite{Richards2007,Mueller2012}, and a more performant and less restrictive version of the consistency constraint based DMPC in \cite{farina2012,farina2014} for dynamically decoupled systems. The preliminary version of the proposed DMPC scheme as presented in \cite{Wiltz2022c} is restricted to nominal dynamics and the pairwise coupling of states through constraints which are limitations that we overcome in this paper.

The remainder is structured as follows. In Sec.~\ref{sec:problem formulation}, we present the partitioned system and the control objective. In Sec.~\ref{sec:main results - dmpc problem}, we define the local optimization problems (Sec.~\ref{subsec:local optimization problems}), present assumptions that allow for their parallelized evaluation and ensure robust asymptotic stability (Sec.~\ref{subsec:consistency constraints}), and derive guarantees of the closed-loop system (Sec.~\ref{subsec:guarantees}). A brief discussion concludes the section. In Sec.~\ref{sec:main results - algorithm}, we provide details on the initialization, the computation of reference trajectories and summarize the overall DMPC algorithm. In Sec.~\ref{sec:simulation}, the algorithm's effectiveness is demonstrated, and in Sec.~\ref{sec:conclusion} some conclusions are drawn.

\subsection{Notation} 
\label{subsec:notation}
A continuous function $ \gamma: \bbR_{+}\rightarrow\bbR_{+} $ is a class $ \calK $ function if it is strictly increasing and $ \gamma(0) = 0 $. If the domain of a trajectory $ x[\kappa] $ with $ \kappa \!=\! a\!,\!\dots\!,\!b $, $ a,b\in\bbN $, is clear from the context, we also write $ x[\cdot] $. By $ x[\cdot|k] $, we denote a trajectory that is computed at time step~$ k $. The short-hand $ x_{i\in\calV} $ is equivalent to $ \lbrace x_i \rbrace_{i\in\calV} $ where $ \calV $ is a set of indices. Let $ \calA, \calB \subset \bbR^n $. Their Minkowski sum is $ \calA \oplus \calB := \lbrace a+b \in \bbR^n \; | \; a \in \calA,\; b \in \calB \rbrace  $, the Pontryagin difference $ \calA \ominus \calB := \lbrace a\in\bbR^n \; | \; a+b\in \calA, \; \forall b\in\calB \rbrace $, and $ \calA \times \calB $ their Cartesian product. $ \bigtimes_{i} $, $ \bigoplus_{i} $ denote the repeated evaluation of the respective operations over sets of indices. The Hausdorff distance $ d_{H} $ is defined as the minimal distance of a point $ x\in\bbR^{n} $ to a set~$ \calA $, i.e., $ d_{H}(x,\calA) := \text{inf}\lbrace ||x-x'|| \, | \, x'\in\calA \rbrace $; $ ||\cdot|| $ denotes the Euclidean norm. Let $ \bm{A}\in\bbR^{m\times n} $ be a matrix, and $ \alpha\in\bbR $ a scalar. Then, $ \alpha\calA := \lbrace y \, | \, y=\alpha x, x \in \calA \rbrace $ and $ \bm{A}\calA := \lbrace y \, | \, y=\bm{A} x, x \in \calA \rbrace $. Finally, $ \mathbf{0}, \mathbf{1} $ denote vectors of all zeros or ones, $ \bm{I} $ an identity matrix, and for $ x\in\bbR^{n} $ and $ \bm{P}\in\bbR^{n\times n} $ positive-definite, we define the weighted norm $ ||x||_{\bm{P}} := \sqrt{x^T \bm{P} x} $.


\section{Preliminaries}
\label{sec:problem formulation}
In this section, we introduce the system dynamics of a partitioned system and its constraints, define the network topology, review robustness related concepts, and finally state the control objective.

\subsection{System Dynamics and Constraints}
\label{subsec:II.A}
Consider a distributed system consisting of subsystems $ i\in\calV = \lbrace 1,\dots,|\calV| \rbrace $ which are dynamically decoupled and behave according to the discrete-time dynamics
\begin{align}
	\label{eq:subsystem dynamics}
	x_i[k+1] = f_{i}(x_i[k],u_i[k],w_i[k]), \qquad x_i[0] = x_{0,i}
\end{align}
where $ x_i\in\bbR^{n_i} $, $ u_i\in\bbR^{m_i} $ denote the actual state and input of subsystem $ i $, respectively, and $ w_i\in\calW_{i} \subset \bbR^{q_i} $ a bounded uncertainty. 
In stack-vector form, the dynamics of the overall system are correspondingly given by
\begin{align}
	\label{eq:system dynamics}
	x[k+1] = f(x[k],u[k],w[k]), \qquad x[0] = x_{0}
\end{align}
where $ f: \bbR^{n}\times \bbR^{m} \times \bbR^{q} \rightarrow \bbR^{n} $ with $ n:=\sum_{i} n_{i} $, $ m:=\sum_{i} m_{i} $, $ q:=\sum_{i} q_{i} $, stack vectors $ x \!=\! [x_1^{T},\dots,x_{|\calV|}^{T}]^{T} $, $ u \!=\! [u_1^{T},\dots,u_{|\calV|}^{T}]^{T} $, $ w \!=\! [w_1^{T},\dots,w_{|\calV|}^{T}]^{T} $, and $ x_{0} \!=\! [x_{0,1}^{T},\dots,x_{0,|\calV|}^{T}]^{T} $. 

All subsystems may be subject to coupled and non-coupled state constraints. If there exists a coupled state constraint of subsystem $ i $ that depends on state $ x_j $ of subsystem $ j $, then we call subsystem~$ j $ a \emph{neighbor} of subsystem~$ i $ and we write $ j\in\calN_i $ where $ \calN_i $ is the set of all neighboring subsystems of $ i $. Let subsystem~$ i $ be subject to $ {R}_i $ coupled state constraints. Then, we define the non-coupled and coupled state constraints of subsystem~$ i $, respectively, via inequalities as
\begin{subequations}
	\label{eq:state constraints}
	\begin{align}
		\label{seq:state constraints state}
		h_i(x_i) &\leq \mathbf{0} \\
		\label{seq:state constraints coupled state}
		c_{i,r}(x_i,x_{j\in\calN_{i,r}}) &\leq \mathbf{0}, \qquad r = 1,\dots,R_i
	\end{align}
\end{subequations}
where $ h_i: \bbR^{n_i} \rightarrow \bbR^{s_i} $ and $ c_{i,r}: \bigtimes_{j\in\lbrace i \rbrace\cup\calN_{i,r}}\bbR^{n_{j}} \!\rightarrow\! \bbR^{s_{i,r}} $ are some continuous functions, and $ \calN_{i,r}\subseteq \calN_{i} $ specifies those neighbors of subsystem $ i $ whose states are coupled with subsystem $ i $ via coupled state constraint $ r $; $ s_i, s_{i,r} \in \bbN_{>0} $ are some constants. 
The corresponding state constraint sets are defined as
\begin{align*}
	\calX_i \!&:=\! \lbrace x_i \in \bbR^{n_i} \, | \, h_i(x_i) \leq \mathbf{0} \rbrace\\
	\calX_{i,r}(x_{j\in\calN_{i,r}}) \!&:=\! \lbrace x_i\in\bbR^{n_i}  |  c_{i,r}(x_i,x_{j\in\calN_{i,r}}) \!\leq\! \mathbf{0} \rbrace
\end{align*}
where $ \calX_{i,r} $ is a set valued function. Moreover, all subsystems $ i\!\in\!\calV $ are subject to input constraints 
\begin{align}
	\label{eq:input constraint}
	u_i\in\calU_i\subseteq\bbR^{m_i}.
\end{align}

From the actual subsystem dynamics \eqref{eq:subsystem dynamics}, we distinguish the \emph{nominal dynamics} of the undisturbed system which are given by
\begin{align}
	\label{eq:nominal subsystem dynamics}
	\begin{split}
		\hat{x}_i[k+1] = \hat{f}_i(\hat{x}_i[k],\hat{u}_i[k]) := f_i(\hat{x}_i[k],\hat{u}_i[k],0), \\ \qquad \hat{x}_i[0]=x_{0,i}
	\end{split}
\end{align}
where $ \hat{x}_i\in\bbR^{n_i} $, $ \hat{u}_i\in\bbR^{m_i} $ denote the nominal state and nominal input of subsystem $ i $, respectively. The nominal dynamics of the overall system are denoted by
\begin{align}
	\label{eq:nominal system dynamics}
	\hat{x}[k+1] = \hat{f}(\hat{x}[k],\hat{u}[k]) := f(\hat{x}[k],\hat{u}[k],0).
\end{align}


\subsection{Network Topology}
The coupled state constraints define a graph structure on the distributed system under consideration. The graph is given as $ \calG = (\calV,\calE) $ where $ \calE := \lbrace (i,j) \, | \, j\in\calN_i \rbrace $ defines the communication links among the subsystems in~$ \calV $. Hence neighboring subsystems can communicate with each other. Throughout the paper, we assume that graph $ \calG $ is undirected.\footnote{A graph $ \calG $ is undirected if $ (i,j)\in\calE $ implies $ (j,i)\in\calE $.}
Moreover, to avoid that subsystems can behave in an adversarial way and ``force'' neighboring subsystems to infeasible states, we assume that neighboring subsystems have those constraints that couple their states in common. This is formally stated next.

\begin{assum}
		\label{ass:common coupled state constraints}
		Let $ c_{i,r}(x_i, x_{j\in\calN_{i,r}}) $, $ r\!\in\!\{1,\dots,R_{i}\} $, be any coupled state constraint of a subsystem~$ i\!\in\!\calV $. Then all neighbors $ j\in\calN_{i,r}\subseteq \calN_i $ are subject to a constraint
		\begin{align*}
			c_{j,r'}(x_j, x_{j'\in\calN_{j,r'}}) \equiv c_{i,r}(x_i,x_{i'\in\calN_{i,r}})
		\end{align*}
		for some $ r'\in\lbrace 1,\dots,R_{j} \rbrace $ where $ \calN_{j,r'} := (\calN_{i,r}\setminus\{j\})\cup\{i\} $ and where $ \equiv $ denotes the equivalence of the functions.
\end{assum}


\subsection{Robust Stability and Tubes}
\label{subsec:robust satbility and tubes}
In order to handle uncertainties $ w_i $ in the dynamics~\eqref{eq:subsystem dynamics}, we resort to a tube-based approach \cite{Rakovic2004,mayne2005}. Here, an auxiliary controller is employed to bound the deviation of the system's actual state to the predicted state into an invariant set.
To this end, we formally define the deviation of an actual state $ x_{i} $ and nominal state $ \hat{x}_{i} $ as $ p_{i} := x_{i} - \hat{x}_{i} $, and the corresponding dynamics of $ p_{i} $ are
\begin{align}
\label{eq:deviation dynamics}
	p_{i}[k+1] = f_{i}(x_{i}[k],u_{i}[k],w_{i}[k])\! -\! \hat{f}_{i}(\hat{x}_{i}[k],\hat{u}_{i}[k])
\end{align}
with $ p_{i}[0] = 0, \; x_{i}[0] = \hat{x}_{i}[0] = x_{0,i} $. Considering a control signal
\begin{align}
	\label{eq:actual control}
	u_{i}[k] = \hat{u}_{i}[k] + K_{i}(x_{i}[k],\hat{x}_{i}[k])
\end{align}
consisting of a nominal input $ \hat{u}_{i} $ and an auxiliary controller $ K_{i}: \calX_{i} \times \calX_{i} \rightarrow \calU_{i} $, we assume that for \eqref{eq:deviation dynamics} controlled by \eqref{eq:actual control} there exists a robust positively invariant (RPI) set $ \calP_{i} $.

\begin{assum}
	\label{ass:existence of RPI set}
	Let the dynamics of $ p_i $ in~\eqref{eq:deviation dynamics} be controlled by \eqref{eq:actual control}. For all $ i\in\calV $, assume that there exists a neighborhood of the origin $ \calP_{i}\subseteq\bbR^{n_{i}} $ such that $ p_{i}[k+1]\in\calP_{i} $ for all $ p_{i}[k]\in\calP_{i} $, $ w_{i}[k]\in\calW_{i} $, and $ \hat{u}_i[k]\in\calU_i $. Such $ \calP_{i} $ is called an RPI set.
\end{assum}

\begin{rem} 
	\label{remark:rpi sets}
	For the construction of RPI sets $ \calP_{i} $, we refer to the rich literature on robust tube-based MPC, see \cite{Rakovic2005,mayne2005,Rakovic2006,rawlings2009,mayne2011,Yu2013,Koehler2021,Berger2022}. For a general nonlinear system, the dynamics of $ p_i $ in~\eqref{eq:deviation dynamics} cannot be written as a function $ \mathfrak{f}(p_{i}[k],w_{i}[k]) $, i.e., $ p_{i}[k+1] = f_{i}(x_{i}[k],u_{i}[k],w_{i}[k]) - \hat{f}_{i}(\hat{x}_{i}[k],\hat{u}_{i}[k]) \neq \mathfrak{f}(p_{i}[k],w_{i}[k]) $. Therefore, it is not always straightforward to find RPI sets $ \calP_{i} $. Instead, most works on the construction of RPI sets focus on particular classes of systems. In \cite{mayne2005}, linear systems of the form $ x[k+1] = \bm{A}x[k] + \bm{B}u[k] $ are considered where $ \bm{A}, \bm{B} $ are real matrices of respective sizes and the pair $ (\bm{A},\bm{B}) $ is assumed to be controllable. In \cite{Rakovic2006}, this is extended to systems with matched nonlinearities of the form
	\begin{align*}
		x[k+1] = \bm{A}x[k]+\bm{B}(g(x[k])u[k]+\varphi(x[k])) + w[k]
	\end{align*}
	where $ g $ is assumed to be invertible for all $ x\in\calX $. For both dynamics, auxiliary controllers can be found such that the dynamics of $ p_i $ take the form $ p[k+1] = \mathfrak{f}(p_i[k],w_i[k]) $. In contrast, the construction in \cite{mayne2011} does not require that the dynamics of $ p_i $ take this form. Other recent approaches improve uncertainty bounds online~\cite{Koehler2021}, or employ the high-gain idea from funnel control~\cite{Berger2022} (the latter however is confined to continuous-time systems). Our focus in this paper is on integrating $ \calP_{i} $ into the proposed consistency constraint based DMPC scheme irrespective of the particular robust MPC method.
\end{rem}

As a consequence of Ass.~\ref{ass:existence of RPI set}, the actual state $ x_i[k] $ stays in a neighborhood $ \calP_i $ of the nominal state $ \hat{x}_{i}[k] $ for all times $ k $ since $ p_i[k] = x_i[k] - \hat{x}_{i}[k] \in \calP_i $, or equivalently 
\begin{align}
	\label{eq:rpi set}
	x_i[k] \in \hat{x}_i[k] \oplus \calP_i, \quad \forall k\geq 0.
\end{align}
Then, we can determine the set of all inputs that $ K_{i} $ possibly takes and define it as
\begin{align}
	\label{eq:Delta U_i}
	\Delta\calU_{i} := \lbrace K_{i}(x_{i},\hat{x}_{i}) \in \calU_{i} \, | \, x_{i},\hat{x}_{i} \!\in\!\calX_{i}, \; x_{i}\!-\!\hat{x}_{i}\!\in\!\calP_{i} \rbrace.
\end{align}
The resulting tightened constraint sets are given as $ \widehat{\calX}_{i} \!:=\! \calX_{i}\!\ominus\!\calP_{i} $, $ \widehat{\calX}_{i,r}(\hat{x}_{j\in\calN_{i,r}}) \!:=\! \lbrace \hat{x}_{i}  | c_{i,r}(x_{i},x_{j\in\calN_{i,r}}) \!\leq\! 0, \;\forall x_i\!\in\!\hat{x}_{i}\!\oplus\!\calP_{i},\;\forall x_{j} \!\in\! \hat{x}_{j}\!\oplus\! \calP_{j}, \; j\!\in\!\calN_{i,r} \rbrace $, and $ \widehat{\calU}_i := \calU_i\ominus\Delta\calU_{i} $.

By suitably choosing a nominal control input $ \hat{u}_{i} $, we aim at guaranteeing the robust asymptotic stabilization of the overall uncertain dynamics~\eqref{eq:system dynamics}.

\begin{defn}[Robust Asymptotic Stability]
	\label{def:robust asymptotic stability}
	Let $ \calR $ be an RPI set for the autonomous discrete-time system $ x[k+1] = g(x[k],w[k]) $ with an equilibrium point $ \xi = g(\xi,0)\in\calR $ and $ x\in\bbR^{n} $, $ w\in\calW\subseteq\bbR^{q} $. The equilibrium $ \xi $ is said to be \emph{robustly stable} if for each  $ \varepsilon>0 $, there exists a $ \delta = \delta(\varepsilon) > 0 $ such that $ d_{H}(x[0],\calR) < \delta \Rightarrow d_{H}(x[k],\calR)< \varepsilon $ for all $ k>0 $ and arbitrary disturbances $ w\in\calW $; $ d_{H} $ denotes the Hausdorff distance. The equilibrium $ \xi $ is \emph{robustly asymptotically stable} if $ \xi $ is robustly stable and $ d_{H}(x[0],\calR) < \delta \Rightarrow \lim_{k\rightarrow\infty} d_{H}(x[k],\calR) = 0 $.
\end{defn}


At last, the following lemma provides useful relations for set-operations, especially in the context of uncertainties.
\begin{lem}[\cite{Schneider2014}, Lem.~3.1.11]
	\label{lem:minkowski relations}
	Let $ \calA, \calB \subset \bbR^{n} $. Then $ (\calA\oplus\calB)\ominus\calB \supseteq \calA $, and $ (\calA\ominus\calB)\oplus\calB \subseteq \calA $.
\end{lem}


\subsection{Control Objective}
\label{subsec:control objective}

Let $ \xi_i = \hat{f}_i(\xi_i,{u}_{\xi_{i}}) $ be a steady state of the nominal dynamics~\eqref{eq:nominal subsystem dynamics} of subsystem $ i $ for a constant nominal input~$ {u}_{\xi_{i}} $. Moreover, denote the stack vector of all steady states by $ \xi = [\xi_1^T,\dots, \xi_{|\calV|}^T]^T $. For all subsystems $ i\in\calV $, the control objective is to robustly asymptotically stabilize desired states $ \xi_{i\in\calV} $ where
\begin{align}
	\label{eq:target state set}
	\begin{split}
		\xi \in \Xi := \lbrace \xi \, | \, &\xi_i = \hat{f}_i(\xi_i,{u}_{\xi_{i}}), \,    \\
		& \hspace{0cm} \xi_{i}\in\widehat{\calX}_{i}, \, \xi_{i}\in\widehat{\calX}_{i,r}(\xi_{j\in\calN_{i,r}}), \, {u}_{\xi_{i}} \in\widehat{\calU}_i, \\ 
		&\forall r = 1,\dots,R_{i}, \;  i\in\calV \rbrace
	\end{split}
\end{align}
satisfies all nominal state constraints (coupled and uncoupled) and nominal input constraints. Note that the tightened constraint sets $ \widehat{\calX}_{i}, \widehat{\calX}_{i,r}, \widehat{\calU}_i $ as previously defined in Sec.~\ref{subsec:robust satbility and tubes} ensure that none of the actual constraints~\eqref{eq:state constraints} and~\eqref{eq:input constraint} are violated in the desired formation due to the uncertainties. 




\section{Distributed MPC Problems}
\label{sec:main results - dmpc problem}

For each subsystem, we now formulate local optimization problems that can be solved in parallel while guaranteeing the satisfaction of all constraints, most notably that of coupled state constraints. In this section, we provide the theoretic foundations of our approach. Implementation details and the overall DMPC algorithm are presented in the subsequent Sec.~\ref{sec:main results - algorithm}. 


\subsection{Local Optimization Problems}
\label{subsec:local optimization problems}

At every time-step~$ k $, a subsystem~$ i $ predicts a nominal state trajectory $ \hat{x}_i[\kappa | k] $ for $ \kappa \!=\! k,\dots,k\!+\!N $ and a corresponding nominal input trajectory $ \hat{u}_i[\kappa | k] $ for $ \kappa \!=\! k,\dots,k+N\!-\!1 $ in accordance with the nominal dynamics~\eqref{eq:nominal subsystem dynamics}. We call $ N $ the prediction horizon. 
As the actual state $ x_{i} $ is uncertain, and it is only known that $ x_i \!\in\! \hat{x}_{i} \!\oplus\! \calP_i $ according to~\eqref{eq:rpi set}, the nominal state trajectories $ \hat{x}_{i\in\calV}[\cdot|k] $ are determined such that the initial state satisfies
\begin{align}
	\label{eq:initial state nominal state trajectory}
	\hat{x}_i[k|k] \in x_i[k] \oplus (-\calP_{i}), \; i\in\calV
\end{align}
where $ -\calP_{i} = -1\calP_{i} $ (cf. notation in Sec.~\ref{subsec:notation}).
In order to allow for parallelization, we introduce the notion of \emph{consistency constraints}. A consistency constraint guarantees that the nominal state trajectory $ \hat{x}_{i} $ stays in a $ \widehat{\calC}_{i} $-neighborhood of a given reference trajectory $ x_{i}^{\text{ref}} $, namely 
\begin{align}
	\label{eq:consistency constraint}
	\hat{x}_i[\kappa | k] \in x_i^{\text{ref}}[\kappa | k] \oplus \widehat{\calC}_i, \quad \forall \kappa=k,\dots,k+N-1
\end{align}
where $ \widehat{\calC}_i\!\subseteq\! \bbR^{n_i} $ is some compact neighborhood of the origin which we call \emph{consistency constraint set}, and $ x_i^{\text{ref}}[\cdot | k] $ a given reference trajectory at time~$ k $. Both $ x_{i}^{\text{ref}} $ and $ \widehat{\calC}_{i} $ are assumed to be known to all neighbors $ j\in\calN_{i} $ of subsystem~$ i $, and are further specified later. While a consistency constraint allows to optimize the nominal state trajectory in a neighborhood of a reference trajectory, it makes the nominal state trajectory predictable to neighbors. Thereby, consistency constraints give rise to parallelization.

\begin{rem}
	Throughout the paper, we employ reference trajectories as an inherent ingredient to consistency constraint based DMPC which is aligned with the terminology used in \cite{farina2012,farina2014}. This should not be mixed up with reference trajectories in the tracking MPC literature. 
\end{rem}

\begin{figure}[tp]
	\centering
	\def\svgwidth{0.65\columnwidth}
\begingroup%
  \makeatletter%
  \providecommand\color[2][]{%
    \errmessage{(Inkscape) Color is used for the text in Inkscape, but the package 'color.sty' is not loaded}%
    \renewcommand\color[2][]{}%
  }%
  \providecommand\transparent[1]{%
    \errmessage{(Inkscape) Transparency is used (non-zero) for the text in Inkscape, but the package 'transparent.sty' is not loaded}%
    \renewcommand\transparent[1]{}%
  }%
  \providecommand\rotatebox[2]{#2}%
  \newcommand*\fsize{\dimexpr\f@size pt\relax}%
  \newcommand*\lineheight[1]{\fontsize{\fsize}{#1\fsize}\selectfont}%
  \ifx\svgwidth\undefined%
    \setlength{\unitlength}{267.5432256bp}%
    \ifx\svgscale\undefined%
      \relax%
    \else%
      \setlength{\unitlength}{\unitlength * \real{\svgscale}}%
    \fi%
  \else%
    \setlength{\unitlength}{\svgwidth}%
  \fi%
  \global\let\svgwidth\undefined%
  \global\let\svgscale\undefined%
  \makeatother%
  \begin{picture}(1,0.84650665)%
    \lineheight{1}%
    \setlength\tabcolsep{0pt}%
    \put(0,0){\includegraphics[width=\unitlength,page=1]{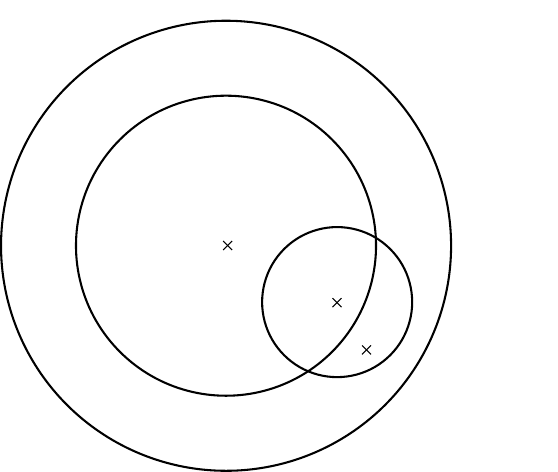}}%
    \put(0.16368216,0.38874866){\color[rgb]{0,0,0}\makebox(0,0)[lt]{\lineheight{0}\smash{\begin{tabular}[t]{l}$x^{\text{ref}}_i[k|k]$\end{tabular}}}}%
    \put(0.24105284,0.24480924){\color[rgb]{0,0,0}\makebox(0,0)[lt]{\lineheight{0}\smash{\begin{tabular}[t]{l}$\hat{x}_i[k|k]$\end{tabular}}}}%
    \put(0.4649419,0.08373621){\color[rgb]{0,0,0}\makebox(0,0)[lt]{\lineheight{0}\smash{\begin{tabular}[t]{l}$x_i[k]$\end{tabular}}}}%
    \put(0,0){\includegraphics[width=\unitlength,page=2]{constraint_sets.pdf}}%
    \put(0.64804128,0.82639003){\color[rgb]{0,0,0}\makebox(0,0)[lt]{\lineheight{0}\smash{\begin{tabular}[t]{l}$\widehat{\calC}_{i}$\end{tabular}}}}%
    \put(0.7698758,0.75761984){\color[rgb]{0,0,0}\makebox(0,0)[lt]{\lineheight{0}\smash{\begin{tabular}[t]{l}$\calC_i$\end{tabular}}}}%
    \put(0.9078236,0.47165857){\color[rgb]{0,0,0}\makebox(0,0)[lt]{\lineheight{0}\smash{\begin{tabular}[t]{l}$\calP_i$\end{tabular}}}}%
    \put(0,0){\includegraphics[width=\unitlength,page=3]{constraint_sets.pdf}}%
  \end{picture}%
\endgroup%

	\caption{Illustration of consistency constraints \eqref{eq:consistency constraint}  (at $ \kappa = k $) and \eqref{eq:actual consistency constraint}. 
	}
	\label{fig:constraint_sets}
\end{figure}

If Ass.~\ref{ass:existence of RPI set} and thereby~\eqref{eq:rpi set} hold, then consistency constraint~\eqref{eq:consistency constraint} also implies
\begin{align}
	\label{eq:actual consistency constraint}
	x_i[k] \in x_i^{\text{ref}}[k | k] \oplus \calC_i \quad \text{with} \quad \calC_{i} := \widehat{\calC}_{i} \oplus \calP_{i}.
\end{align}
Figure~\ref{fig:constraint_sets} illustrates the concept of consistency constraints.

Given $ x_{i}[k] $ and $ x_i^{\text{ref}}[\cdot|k] $, we define the local optimization problem of subsystem~$ i $ at time-step $ k $ as
\begin{align}
	\label{eq:opt criterion}
	J_i^{\ast}(x_i[k]) = \min_{\substack{\hat{x}_{i}[k|k],\hat{u}_i[\cdot | k] 
		} }
	J_i(\hat{x}_i[\cdot | k], \hat{u}_i[\cdot | k]) 
\end{align}
where
\begin{align*}
	J_i(\hat{x}_i[\cdot | k], \hat{u}_i[\cdot | k]) &= \sum_{\kappa = k}^{k+N-1} l_i(\hat{x}_i[\kappa|k],\hat{u}_i[\kappa|k]) \\&\qquad+ {J}_i^f(\hat{x}_i[k+N|k]) \\
	l_i(\hat{x}_i[\kappa|k],\hat{u}_i[\kappa|k]) &= ||\hat{x}_i[\kappa | k] \!-\! \xi_i||^{2}_{\bm{Q}_i} \!+\! ||\hat{u}_i[\kappa|k] \!-\! {u}_{\xi_i}||^{2}_{\bm{R}_i}
\end{align*}
with stage-cost-function $ l_i $, a positive-definite terminal cost-function $ {J}_i^f $, and positive-definite matrices $ \bm{Q}_i, \bm{R}_i $. The local optimality criterion \eqref{eq:opt criterion} is subject to
\begin{subequations}
	\label{eq:opt constraints}
	\begin{align}
		\label{seq:opt constraints 1a}
		\hat{x}_i[k|k] &\;\in x_i[k] \oplus (-\calP_{i}) \\
 		\label{seq:opt constraints 1b}
		\hat{x}_i[\kappa+1|k] &= \hat{f}_i(\hat{x}_i[\kappa|k],\hat{u}_i[\kappa|k])
		 \\
		\label{seq:opt constraints 2}
		\hat{x}_i[\kappa|k] &\in x^{\text{ref}}_{i}[\kappa|k]\oplus \widehat{\calC}_i \\
		\label{seq:opt constraints 3}
		\hat{u}_i[\kappa|k] &\in \widehat{\calU}_i \\
		\label{seq:opt constraints 4}
		\hat{x}_i[k+N|k]&\in\widehat{\calX}_i^{f}
	\end{align}
\end{subequations}
for $ \kappa=k,\dots,k+N-1 $ and some terminal set $ \widehat{\calX}^{f}_i $ to be further specified below. 
We denote those trajectories $ \hat{x}_i $, $ \hat{u}_{i} $ that minimize \eqref{eq:opt criterion} by $ \hat{x}_i^{\ast}[\cdot|k] $ and $ \hat{u}_i^{\ast}[\cdot|k] $. 
Then, the control law for the $ i $-th subsystem~\eqref{eq:subsystem dynamics}, for all $ k\geq0 $, is given by
\begin{align}
	\label{eq:actual input law}
	u^{\text{MPC}}_{i}(x_{i}[k]) := \hat{u}^{\text{MPC}}_{i}(x_{i}[k]) \!+\! K_{i}(x_{i}[k],\hat{x}^{\ast}_{i}[k|k])
\end{align}
where $ \hat{u}^{\text{MPC}}_i(x_i[k]) := \hat{u}^{\ast}_i[k|k] $ and $ K_{i} $ is the auxiliary controller rendering~$ \calP_{i} $ an RPI set (cf. Ass.~\ref{ass:existence of RPI set}).

We impose the following slightly modified standard assumptions \cite{mayne2000} on terminal sets $ \widehat{\calX}_i^{f} $ and terminal cost $ J_i^{f} $.
\begin{assum}[Terminal Constraints]
	\label{ass:terminal set}
	For the terminal sets $ \widehat{\calX}_i^{f} $, the terminal cost function $ J_i^{f}: \widehat{\calX}_i^{f} \rightarrow \bbR_{\geq 0} $, and some state-feedback controller $ k^f_i(x_i) $, $ i\!\in\!\calV $, we assume:
	\begin{enumerate}[leftmargin=*, start=1,label={A\ref{ass:terminal set}.\arabic*},ref={\ref{ass:terminal set}.\arabic*}]
		\item 
		\label{ass:terminal set a1}
		for all $ \widehat{\calX}_i^{f} $, $ i\in\calV $,  
		\begin{enumerate}[label=(\roman*)]
			\item $ \widehat{\calX}_i^{f} \subseteq \calX_i \ominus 
			\calC_{i} $, and
			\item for all $ r=1,\dots,R_{i} $, it holds $ c_{i,r}(x_i,x_{j\in\calN_{i,r}}) \leq \mathbf{0} $ for all $ x_i\in\widehat{\calX}_i^{f}\oplus\calC_{i} $ and for all $ x_j \in\widehat{\calX}_j^{f}\oplus\calC_{j} $, $ j\in\calN_{i,r} $
		\end{enumerate}
		(terminal states satisfy uncoupled~(i) and coupled state constraints~(ii) in a $ \calC_{i} $-neighborhood);
		
		\item 
		\label{ass:terminal set a2}
		$ k^f_{i}(x_i)\in\widehat{\calU}_i $ for all $ x_i\in\widehat{\calX}_i^{f} $ (input constraint satisfaction);
		
		\item 
		\label{ass:terminal set a3}
		$ \hat{f}_i(x_i,k^f_{i}(x_i)) \in\widehat{\calX}_i^{f} $  for all $ x_i\in\widehat{\calX}_i^{f} $ (set invariance);
		
		\item 
		\label{ass:terminal set a4}
		$ J_i^{f}(\hat{f}(x_i,k^f_{i}(x_i))) + l_i(x_i,k^f_i(x_i)) \leq J_i^{f}(x_i) $ for all $ x_i\in\widehat{\calX}_i^{f} $ (terminal cost function is a local Lyapunov function).
	\end{enumerate}
\end{assum}


\subsection{Reference Trajectories and Consistency Constraints}
\label{subsec:consistency constraints}

The role of the consistency constraints is twofold: on the one hand, consistency constraints shall ensure the satisfaction of all state constraints (both coupled and uncoupled) by bounding the predicted nominal state trajectory $ \hat{x}_{i} $ in the neighborhood of its reference trajectory~$ x_{i}^{\text{ref}} $. On the other hand, consistency constraints shall only restrict the evolution of nominal state trajectories in so far such that the feasibility of the local optimization problem \eqref{eq:opt criterion}-\eqref{eq:opt constraints} is preserved. To this end, we assume the following properties of reference trajectories $ x_{i}^{\text{ref}} $, $ i\in\calV $, for given consistency constraint sets~$ \calC_{i} $, $ i\in\calV $.

\begin{assum}[Reference Trajectories]
	\label{ass:consistency constraint}
	For $ \kappa \!=\! k,\dots,k\!+\!N\!-\!1 $, $ k\!\geq\! 0 $, and reference states $ x^{\text{ref}}_i[\kappa | k] $, $ i\!\in\!\calV $, 
	\begin{enumerate}[leftmargin=*, start=1,label={A\ref{ass:consistency constraint}.\arabic*},ref={\ref{ass:consistency constraint}.\arabic*}]
		\item
		\label{ass:consistency constraint a1}
		it holds that
		\begin{align}
			\label{eq:ref trajectory state constraint}
			h_i(x_i) \leq \mathbf{0} \quad \forall x_i \in x^{\text{ref}}_i[\kappa|k]\oplus\calC_i
		\end{align}
		and for all $ r=1,\dots,R_{i}, $
		\begin{align}
			\label{eq:ref trajectory coupled state constraint}
			\begin{split}
				c_{i,r}(x_i,x_{j\in\calN_{i,r}}) \leq \mathbf{0} \quad &\forall x_{i} \in x^{\text{ref}}_{i}[\kappa|k]\oplus\calC_{i}, \\
				&\hspace{-1.5cm} \forall x_{j} \in x^{\text{ref}}_{j}[\kappa|k]\oplus\calC_{j}, \; j\in\calN_{i,r};
			\end{split}
		\end{align}
		
		\item 
		\label{ass:consistency constraint a2}
		and additionally, if $ k>0 $, it holds that
		\begin{align}
			\label{eq:ref trajectory recursive feasibillity constraint}
			\hat{x}^{\ast}_i[\kappa | k-1] &\in x^{\text{ref}}_i[\kappa|k ]\oplus\widehat{\calC}_i.
		\end{align}
	\end{enumerate}
\end{assum}

The first part of Ass.~\ref{ass:consistency constraint} states that consistency constraints imply the satisfaction of state constraints~\eqref{eq:state constraints}. The second part of Ass.~\ref{ass:consistency constraint} leads to recursive feasibility. In Sec.~\ref{subsec:detremination of reference states}, we propose a method for the determination of reference trajectories that satisfy Ass.~\ref{ass:consistency constraint}.

At time~$ k=0 $, we call a set of reference trajectories $ {x}^{\text{ref}}_i[\kappa|0] $, $ \kappa=0,\dots,N $, $ i\in\calV $, and the corresponding input trajectories $ {u}^{\text{ref}}_i[\kappa|0] $, $ \kappa=0,\dots,N-1 $, \emph{initially feasible} if 
\begin{subequations}
	\label{eq:opt init trajectories}
	\begin{align}
		\label{seq:opt init trajectories 1}
		\begin{split}
			{x}^{\text{ref}}_i[0|0] &= x_{0,i}\\
			{x}^{\text{ref}}_i[\kappa+1|0] &= \hat{f}_i({x}^{\text{ref}}_i[\kappa|0],{u}^{\text{ref}}_i[\kappa|0])
		\end{split}\\
		\label{seq:opt init trajectories 2}
		h_i(x_i)&\leq \mathbf{0} \quad \forall x_i \in x^{\text{ref}}_i[\kappa|0] \oplus \calC_{i}\\
		\label{seq:opt init trajectories 3}
		\begin{split}
			c_{i,r}(x_i,x_{j\in\calN_{i,r}}) &\leq \mathbf{0} \quad \forall x_i \in x^{\text{ref}}_i[\kappa|0]\oplus \calC_{i}\\
			&\hspace{-0.5cm}\forall x_{j} \in x^{\text{ref}}_{j}[\kappa|0]\oplus\calC_{j}, \; j\in\calN_{i,r}\\
			&\hspace{-0.5cm} \forall r \in \lbrace 1,\dots,R_{i} \rbrace
		\end{split}\\
		\label{seq:opt init trajectories 4}
		{u}^{\text{ref}}_i[\kappa|0] &\in\widehat{\calU}_i\\
		\label{seq:opt init trajectories 5}
		{x}^{\text{ref}}_i[N|0] &\in \widehat{\calX}_i^{f}
	\end{align}
\end{subequations}
where $ \widehat{\calX}_{i}^{f} $ satisfies Ass.~\ref{ass:terminal set}.


\subsection{Guarantees of the Distributed MPC Problems}
\label{subsec:guarantees}

Now, we show that the local optimization problems always remain feasible given that there exist initially feasible trajectories, and that the closed-loop system is robust asymptotically stable with respect to $ \xi_{i} $, $ i\in\calV $. The dynamics of the closed-loop system resulting from the application of $ u_{i}^{\text{MPC}} $ defined in~\eqref{eq:actual input law} are given~as
\begin{align}
	\label{eq:closed loop dynamics}
	x_i[k+1] &= f_{i}(x_{i}[k],u^{\text{MPC}}_{i}(x_{i}[k]),w_{i}[k]), \quad\! i \in \calV.
\end{align}

\begin{thm}
	\label{theorem:main theorem}
	Let Ass.~\ref{ass:existence of RPI set},~\ref{ass:terminal set} and~\ref{ass:consistency constraint} hold, and let there be a set of initially feasible reference trajectories $ x^{\text{ref}}_{i}[\cdot|0] $, $ i\in\calV $. Then for each subsystem $ i\in\calV $, the local optimization problems~\eqref{eq:opt criterion} subject to~\eqref{eq:opt constraints} are recursively feasible with respect to the closed-loop dynamics~\eqref{eq:closed loop dynamics} for any $ w_{i}[k]\in\calW_{i} $. Moreover, the closed-loop dynamics~\eqref{eq:closed loop dynamics} are robustly asymptotically stable with respect to~$ \xi_{i} $,~$ i\in\calV $. 
\end{thm}
\begin{pf}
	In a first step, we prove recursive feasibility. Thereafter robust asymptotic stability is proven.
	
	\emph{Recursive Feasibility:} We have to show that there exist feasible solutions to the local optimization problems~\eqref{eq:opt criterion} subject to~\eqref{eq:opt constraints} for $ k>0 $ given that there exist initially feasible reference trajectories at $ k=0 $. To this end, we recursively construct candidate trajectories $ \hat{x}^{\text{c}}_i[\kappa|k] $ for $ \kappa=k,\dots,k+N $ and $ \hat{u}^{\text{c}}_i[\kappa|k] $ for $ \kappa=k,\dots,k+N-1 $ that satisfy constraints~\eqref{eq:opt constraints}. 
	
	Firstly, consider $ k\!=\!0 $ and choose candidate trajectories
	\begin{subequations}
		\begin{align*}
			\hat{x}_i^{\text{c}}[\kappa|0] &= x^{\text{ref}}_i[\kappa|0] \quad \text{for } \kappa = 0,\dots,N, \\
			\hat{u}_i^{\text{c}}[\kappa|0] &= u^{\text{ref}}_i[\kappa|0] \quad \text{for } \kappa = 0,\dots,N-1
		\end{align*}
	\end{subequations}
	for all $ i\in\calV $ where $ x^{\text{ref}}_i[\cdot|0] $ denotes an initially feasible reference trajectory and $ u^{\text{ref}}_i[\cdot|0] $ its corresponding input trajectory which together satisfy~\eqref{eq:opt init trajectories}. Now, we show that $ \hat{x}_i^{\text{c}}[\cdot|0] $ and $ \hat{u}_i^{\text{c}}[\cdot|0] $ also satisfy~\eqref{eq:opt constraints}.
	
	Since $ x^{\text{ref}}_i[\cdot|0], u^{\text{ref}}_i[\cdot|0] $ satisfy \eqref{seq:opt init trajectories 1}, it holds that 
	\begin{align*}
		\hat{x}_i^{\text{c}}[0|0] = x^{\text{ref}}_i[0|0] \stackrel{\eqref{seq:opt init trajectories 1}}{=} x_{0,i} = x_{i}[0]
	\end{align*}
	and it follows that \eqref{seq:opt constraints 1a} and \eqref{seq:opt constraints 1b} are satisfied by $ \hat{x}_i^{\text{c}}[\cdot|0], \hat{u}_i^{\text{c}}[\cdot|0] $ for all $ i\!\in\!\calV $.
	Because $ \hat{x}_i^{\text{c}}[\cdot|0] = x_i^{\text{ref}}[\cdot|0] $ and $ \calC_i $ is a closed neighborhood of the origin, it follows that \eqref{seq:opt constraints 2} is also satisfied by $ \hat{x}_i^{\text{c}}[\cdot|0] $. The satisfaction of \eqref{seq:opt constraints 3}-\eqref{seq:opt constraints 4} trivially follows from \eqref{seq:opt init trajectories 4}-\eqref{seq:opt init trajectories 5}. Thereby, we have shown that there exist feasible solutions to the local optimization problem \eqref{eq:opt criterion}-\eqref{eq:opt constraints} at $ k\!=\!0 $. 
	
	In a next step, we show that there exist feasible solutions for $ k\!>\!0 $ using induction. Assume that for all $ i\in\calV $ at time-step~$ k $, there exist predicted nominal state and input trajectories, namely $ \hat{x}^{\ast}_i[\cdot|k] $ and $ \hat{u}^{\ast}_i[\cdot|k] $, that solve~\eqref{eq:opt criterion} subject to constraints~\eqref{eq:opt constraints} (induction hypothesis). 
	As we have shown in the previous paragraph, the induction hypothesis is initially fulfilled at $ k=0 $.
	We now show that given the induction hypothesis holds, there also exist feasible candidate trajectories $ \hat{x}_i^{\text{c}}[\cdot|k+1] $, $ \hat{u}_i^{\text{c}}[\cdot|k+1] $ for all $ i\in\calV $ at time $ k+1 $ (induction step). Therefore, we construct for time-steps $ k+1 $, $ k\geq0 $, and all subsystems $ i\in\calV $ candidate trajectories
	\begin{subequations}
		\label{eq:candiate trajectories}
		\begin{align}
		\label{seq:candiate trajectories states}
		\hat{x}_i^{\text{c}}[\kappa|k\!+\!1] &= 
		\begin{cases}
		\hat{x}^{\ast}_i[\kappa|k] & \hspace{-3.3cm}\text{for } \kappa = k\!+\!1,\dots,k\!+\!N \\
		\hat{f}_i(\hat{x}^{\ast}_i[k\!+\!N|k],k_i^{f}(\hat{x}^{\ast}_i[k\!+\!N|k])) & \\&\hspace{-3.3cm}\text{for } \kappa = k\!+\!N\!+\!1
		\end{cases}\\
		\label{seq:candiate trajectories inputs}
		\hat{u}_i^{\text{c}}[\kappa|k\!+\!1] &= 
		\begin{cases}
		\hat{u}^{\ast}_i[\kappa|k] & \hspace{-0.2cm}\text{for } \kappa \!=\! k\!\!+\!\!1,\dots,k\!\!+\!\!N\!\!-\!\!1 \\
		k_i^{f}(\hat{x}_i^{\ast}[k\!\!+\!\!N|k]) & \hspace{-0.2cm} \text{for } \kappa \!=\! k\!\!+\!\!N
		\end{cases}
		\end{align}
	\end{subequations}
	where $ k_i^{f} $ denotes the auxiliary controller associated with $ \widehat{\calX}_{i}^{f} $ and $ J_{i}^{f} $ as specified in Ass.~\ref{ass:terminal set}. Now, we show that $ \hat{x}_i^{\text{c}}[\cdot|k \! +\!1] $ and $ \hat{u}_i^{\text{c}}[\cdot|k \! +\!1] $ satisfy~\eqref{eq:opt constraints}.
	At first, we obtain for the initial value of $ \hat{x}^{\text{c}}_i[\cdot|k\!+\!1] $
	\begin{align*}
		&\hat{x}_i^{\text{c}}[k\!+\!1|k\!+\!1] = \hat{x}^{\ast}_i[k\!+\!1|k] = \hat{f}_i(\hat{x}^{\ast}_i[k|k],\hat{u}^{\ast}_i[k|k]) \\ 
		&\quad= \hat{f}_i(\hat{x}^{\ast}_i[k|k],\hat{u}^{\text{MPC}}_i(x_{i}[k]))\\
		&\quad\stackrel{\eqref{eq:deviation dynamics}}{=} f_{i}(x_{i}[k], u^{\text{MPC}}_i(x_{i}[k]),w_i[k]) \!-\! p_{i}[k\!+\!1] \\
		&\quad \stackrel{\eqref{eq:subsystem dynamics}}{=} x_{i}[k\!+\!1] - p_{i}[k\!+\!1] \in x_{i}[k\!+\!1] \oplus (-\calP_{i})
	\end{align*}
	 and \eqref{seq:opt constraints 1a} is satisfied. As $ \hat{x}^{\ast}_i[\cdot|k] $, $ \hat{u}^{\ast}_i[\cdot|k] $ satisfy~\eqref{seq:opt constraints 1b} according to the induction assumption, candidate trajectories~\eqref{eq:candiate trajectories} satisfy~\eqref{seq:opt constraints 1b} as well. Due to Ass.~\ref{ass:consistency constraint a2}, the candidate trajectory $ \hat{x}_i^{\text{c}}[\kappa|k\!+\!1] = \hat{x}_i^{\ast}[\kappa|k\!+\!1] $, $ \kappa = k,\dots,k\!+\!N\!-\!1 $, satisfies \eqref{seq:opt constraints 2} for all $ i\in\calV $. 
	Besides, as $ \hat{u}_i^{\text{c}}[\kappa|k\!+\!1] = \hat{u}_i^{\ast}[\kappa|k] $, \eqref{seq:opt constraints 3} is trivially satisfied for $ \kappa = k\!+\!1,\dots,k\!+\!N\!-\!1 $. Moreover, $ \hat{u}_i^{\text{c}}[k\!+\!N|k\!+\!1] = k_i^{f}(\hat{x}_i^{\ast}[k\!+\!N|k])\in\widehat{\calU}_i $ holds due to Ass.~\ref{ass:terminal set a2} which applies because $ \hat{x}_i^{\ast}[k\!+\!N|k]\in\widehat{\calX}_i^{f} $, and the satisfaction of \eqref{seq:opt constraints 3} also follows for $ \kappa = k\!+\!N $. 	
	Finally, it follows from Ass.~\ref{ass:terminal set a3} that
	\begin{align*}
		\hat{x}_i^{\text{c}}[k\!\! +\!\! N\!\!+\!\!1|k\!\!+\!\!1] \!=\! \hat{f}_i(\hat{x}^{\ast}_i[k\! +\! N|k],k_{i}^{f}(\hat{x}^{\ast}_i[k\!+\!N|k]))\!\in\!\widehat{\calX}_{i}^{f}.
	\end{align*}
	Thus, \eqref{seq:opt constraints 4} is satisfied as well.
	Now, as the induction hypothesis is also initially fulfilled, it inductively follows that for all $ k>0 $ there exists a feasible solution to the optimization problems \eqref{eq:opt criterion}-\eqref{eq:opt constraints}, and we conclude recursive feasibility.
	
	\emph{Robust asymptotic stability:}	At first, we define the auxiliary cost function
	\begin{align}
		\label{eq:nominal cost function}
		V_{i}(x_i[k]) := \min_{\substack{\hat{u}_i[\cdot | k] }} J_i(\hat{x}_i[\cdot | k], \hat{u}_i[\cdot | k])
	\end{align}
	subject to $ \hat{x}_i[k|k]=x_i[k] $ (which implies~\eqref{seq:opt constraints 1a}) and \eqref{seq:opt constraints 1b}--\eqref{seq:opt constraints 4}; $ J_{i} $ is the same as in \eqref{eq:opt criterion}. 
	Then, it holds
	\begin{align}
		\label{eq:suboptimality}
		\begin{split}
			J^{\ast}_{i}(x_i[k\!+\!1]) \!=\! \hspace{-0.2cm}\min_{z\in x_i[k\!+\!1] \oplus (-\calP_{i})}\hspace{-0.3cm} V_{i}(z) \!\leq\! V_i(\hat{x}^{\ast}_i[k\!+\!1|k]).
		\end{split}
	\end{align}
	Note that here $ \hat{x}^{\ast}_i[k\!+\!1|k] \!\in\! x_{i}[k\!+\!1] \!\oplus\! (-\calP_{i}) $, and hence the inequality follows from the suboptimality of $ \hat{x}^{\ast}_i[k+1|k] $ with respect to the minimization. Using this preliminary result, we show in the remainder of the proof that $ J_{i}^{\ast} $ is a Lyapunov function in terms of \cite[Thm.~1]{Kalman1960a}.
	
	By the positive definiteness of $ \bm{Q}_i, \bm{R}_i $, it holds for the stage cost function $ l_{i}(\xi_{i},u_{\xi_{i}})=0 $ and $ l_{i}(\hat{x}_{i},\hat{u}_{i}) > 0 $ for all $ \hat{x}_{i} \neq \xi_{i}$, $ \hat{u}_{i} \neq u_{\xi_{i}} $. Thus, there exists a class $ \calK $ function $ \alpha $ such that $ \alpha(0)=0 $ and $ 0<\alpha(d_{H}(x_{i},\xi_{i}\oplus\calP_{i}))\leq J_{i}^{\ast}(x_{i}) $ for all $ x_{i}\notin\xi\oplus\calP_{i} $. This is because
	\begin{align*}
		J_{i}^{\ast}(x_{i}) &\geq \min_{\substack{\hat{x}_{i}[k|k]\in x_{i}\oplus (-\calP_{i}) \\ \hat{u}_{i}[k|k]\in\calU_{i}}} l_{i}(\hat{x}_{i}[k|k],\hat{u}_{i}[k|k]) \\
		&\geq \min_{\hat{x}_{i}\in x_{i}\oplus (-\calP_{i})} ||\hat{x}_{i} -\xi_{i}||_{\bm{Q}_{i}} \\ 
		&\geq \sqrt{\lambda_{\text{min}}(\bm{Q}_{i})} \min_{\hat{x}_{i}\in x_{i}\oplus (-\calP_{i})} ||\hat{x}_{i}-\xi_{i}|| \\
		&= \sqrt{\lambda_{\text{min}}(\bm{Q}_{i})}\; d_{H}(x_{i}, \xi_{i}\oplus \calP_{i}) 
	\end{align*}
	and we identify $ \alpha(z)=\sqrt{\lambda_{\text{min}}(\bm{Q}_{i})} z $ where $ \lambda_{\text{min}}(\bm{Q}_{i}) $ denotes the smallest eigenvalue of $ \bm{Q}_{i} $. Moreover, for $ x_{i}[k]\in \xi_{i}\oplus\calP_{i} $, the unique optimal solution to the local optimization problem~\eqref{eq:opt criterion}-\eqref{eq:opt constraints} is $ \hat{x}^{\ast}_{i}[k|k] \!=\! \xi_{i} $, $ \hat{x}_{i}^{\ast}[\cdot|k] \!\equiv\! \xi_{i} $, $ \hat{u}_{i}^{\ast}[\cdot|k] \!\equiv\! u_{\xi_{i}} $. Therefore, $ J_{i}^{\ast}(x_{i})\!=\!0 $ for all $ x_{i}[k]\!\in\! \xi_{i}\!\oplus\!\calP_{i} $. Thus, by the feasibility of~\eqref{eq:opt criterion}-\eqref{eq:opt constraints} as shown in the recursive feasibility proof, there exists a class~$ \calK $ function $ \beta $ such that $ J_{i}^{\ast}(x_{i}) \leq \beta(d_{H}(x_{i},\xi_{i}\oplus\calP_{i})) $.
	
	Next, we investigate the descent on $ J_{i}^{\ast} $ for the closed-loop system. At first, we observe
	\begin{align}
		&J_i^{\ast}(x_i[k\!+\!1])\!-\!J_i^{\ast}(x_i[k]) \stackrel{\eqref{eq:suboptimality}}{\leq} V_i(\hat{x}^{\ast}_i[k\!+\!1|k])\!-\!J_i^{\ast}(x_i[k])\nonumber\\
		\label{eq:mainthm V1}
		&\qquad \leq J_i(\hat{x}_i^{\text{c}}[\cdot|k+1],\hat{u}_i^{\text{c}}[\cdot|k+1]) - J_i^{\ast}(x_i[k])
	\end{align}
	where the latter inequality follows from definition~\eqref{eq:nominal cost function} and the suboptimality of the candidate trajectories. From this, we further derive
	\begin{align}
		&J_i(\hat{x}_i^{\text{c}}[\cdot|k+1],\hat{u}_i^{\text{c}}[\cdot|k+1]) - J_i^{\ast}(x_i[k]) \nonumber\\
		& = \!\!\!\!\!\sum_{\kappa=k+1}^{k+N}\!\!\!\! l_i(\hat{x}_i^{\text{c}}[\kappa|k\!\!+\!\!1],\hat{u}_i^{\text{c}}[\kappa|k\!\!+\!\!1]) \!+\! J_i^{f}(\hat{x}_i^{\text{c}}[k\!\!+\!\!N\!\!+\!\!1|k\!\!+\!\!1]) \nonumber\\
		& \quad -\sum_{\kappa=k}^{k+N-1} l_i(\hat{x}_i^{\ast}[\kappa|k],\hat{u}_i^{\ast}[\kappa|k]) - J_i^{f}(\hat{x}_i^{\ast}[k+N|k]) \nonumber\\
		& \stackrel{\text{\eqref{eq:candiate trajectories}}}{=} \!\! l_i(\hat{x}_i^{\ast}\![k\!\!+\!\!N|k],k_{i}^{f}(\hat{x}_i^{\ast}\![k\!\!+\!\!N|k])) \!-\! l_i(\hat{x}^{\ast}_i\![k|k],\hat{u}_i^{\ast}\![k|k]) \nonumber\\
		&\;\, +\! J_i^{f}\!(\hat{f}_i(\hat{x}_i^{\ast}\![k\!\!+\!\!N|k],k_{i}^{f}(\hat{x}_i^{\ast}\![k\!\!+\!\!N|k]))) \!-\! J_i^{f}\!(\hat{x}_i^{\ast}\![k\!\!+\!\!N|k])\nonumber\\
		&\!\!\!\stackrel{\text{Ass.~\ref{ass:terminal set a4}}}{\leq} - l_i(\hat{x}^{\ast}_i[k|k],\hat{u}_i^{\ast}[k|k]) \nonumber\\
		\label{eq:mainthm V2}
		&\;\,\leq -||\hat{x}^{\ast}_i[k|k] \!-\! \xi_i||^{2}_{\bm{Q}_i} = -\gamma_{V_i}(||\hat{x}^{\ast}_{i}[k|k]\!-\!\xi_{i}||)
	\end{align}
	where $ \gamma_{V_i} $ is a class $ \calK $ function. Note that 	
	\begin{align*}
		\begin{split}
			&|| \hat{x}^{\ast}_{i}[k|k]\!-\!\xi_{i} || = d_{H}(\hat{x}^{\ast}_{i}[k|k]\!+\! p_{i}[k|k], \xi_{i}\!+\!p_{i}[k|k]) \\
			&\quad\geq d_{H}(\hat{x}^{\ast}_{i}[k|k]\!+\!p_{i}[k|k],\xi_{i}\!\oplus\!\calP_{i}) = d_{H}(x_{i}[k],\xi_{i}\!\oplus\!\calP_{i})
		\end{split}
	\end{align*}
	where $ p_{i}[k|k] = x_{i}[k] - \hat{x}^{\ast}_{i}[k|k]\in\calP_{i} $, which finally yields, together with~\eqref{eq:mainthm V1}-\eqref{eq:mainthm V2},
	\begin{align*}
		\begin{split}
			J_i^{\ast}(x_i[k\!+\!1])\!-\!J_i^{\ast}(x_i[k]) \!\leq\! -\gamma_{V_i}(d_{H}(x_{i}[k],\xi_{i}\!\oplus\!\calP_{i})) <0
		\end{split}
	\end{align*}
	for all $ x_{i}[k]\notin \xi_{i}\oplus\calP_{i} $. Thus, by \cite[Thm.~1]{Kalman1960a}, we have shown that $ J_{i}^{\ast} $ is a Lyapunov function. We further conclude the asymptotic stability of $ \xi_{i}\oplus\calP_{i} $ under closed-loop dynamics~\eqref{eq:closed loop dynamics}, and equivalently the robust asymptotic stability of $ \xi_{i} $ for all $ i\in\calV $.  {\color{white}.}\hfill$ \Box $
\end{pf}

Next, we show that state and input trajectories of the closed-loop system~\eqref{eq:closed loop dynamics} satisfy all constraints.

\begin{prop}
		\label{lemma:constraint satisfaction}
		Let the same premises hold as in Thm.~\ref{theorem:main theorem}, i.e., let Ass.~\ref{ass:existence of RPI set},~\ref{ass:terminal set} and~\ref{ass:consistency constraint} hold, and let there be a set of initially feasible reference trajectories $ x^{\text{ref}}_{i\in\calV}[\cdot|0] $. Then for all $ i\in\calV $, the state trajectory $ x_{i}[k] $, $ k\geq0 $, of the closed-loop system~\eqref{eq:closed loop dynamics} and the corresponding input trajectory $ u_{i}[k] = u^{\text{MPC}}_{i}(x_{i}[k]) $, $ k\geq0 $, satisfy state constraints~\eqref{eq:state constraints} and input constraints~\eqref{eq:input constraint}. 
\end{prop}

\begin{pf}
	From Thm.~\ref{theorem:main theorem}, it follows that the local optimization problems are recursively feasible and hence state and input trajectories $ x_{i}[k] $ and $ u_{i}[k] $ of the closed-loop dynamics~\eqref{eq:closed loop dynamics} are well-defined for $ k\geq0 $. According to~\eqref{eq:actual input law}, the input trajectory is given by $ u_{i}[k]=u^{\text{MPC}}_i(x_i[k]) = \hat{u}^{\ast}_{i}[k|k] + K_{i}(x_{i}[k],\hat{x}^{\ast}_{i}[k|k]) $ where $ \hat{u}^{\ast}_{i}[k|k] $ is such that \eqref{seq:opt constraints 3} holds, and by definition~\eqref{eq:Delta U_i}, we have $ K_{i}(\cdot,\cdot)\in\Delta\calU_{i} $. Then, it follows that  
	\begin{align*}
		u_{i}[k]\in\widehat{\calU}_{i} \oplus \Delta\calU_{i} = (\calU_{i}\ominus\Delta\calU_{i})\oplus\Delta\calU_{i} \stackrel{\text{Lem.~\ref{lem:minkowski relations}}}{\subseteq} \calU_{i},
	\end{align*}
	and input constraint~\eqref{eq:input constraint} holds. Next, due to consistency constraint \eqref{seq:opt constraints 2} and $ p_i[k]=x_{i}[k]-\hat{x}_{i}[k]\in\calP_{i} $ (cf.~\eqref{eq:rpi set}), it follows that for all $ i\in\calV $
	\begin{align*}
		x_{i}[k] \!=\! \hat{x}_{i}[k] \!+\! p_{i}[k] \!\in\! (x_{i}^{\text{ref}}[k|k] \!\oplus\!\widehat{\calC}_{i})\oplus\calP_{i} \!=\! x_{i}^{\text{ref}}[k|k] \!\oplus\! \calC_{i}.
	\end{align*}
	Then due to Ass.~\ref{ass:consistency constraint}, the satisfaction of state constraints \eqref{seq:state constraints state} and coupled state constraints \eqref{seq:state constraints coupled state} follows from \eqref{eq:ref trajectory state constraint}-\eqref{eq:ref trajectory coupled state constraint}. {\color{white}.}\hfill$ \Box $
\end{pf}

\subsection{Discussion}
\label{sec:discussion}
If reference trajectories~$ x_{i}^{\text{ref}} $ and consistency constraint sets~$ \calC_{i} $ satisfy Ass.~\ref{ass:consistency constraint a1}, then state constraint satisfaction is implied via consistency constraint~\eqref{seq:opt constraints 3}. Therefore, no further state constraints apart from the consistency constraint need to be considered in the local optimization problem \eqref{eq:opt criterion}-\eqref{eq:opt constraints}. Hence, the local optimization problems are subject to fewer constraints compared to other DMPC schemes allowing for coupled state constraints, see \cite{Nikou2019,Richards2007,Trodden2010,farina2012}. 

Additionally, if the consistency constraint sets, namely~$ \widehat{\calC}_{i} $, are chosen to be convex, then the local optimization problems~\eqref{eq:opt criterion}-\eqref{eq:opt constraints} may be convex even if the original state constraints~\eqref{eq:state constraints} are non-convex. In particular, this is the case if dynamics~\eqref{eq:subsystem dynamics} are linear and the input constraint sets $ \calU_{i} $ convex; note that linear dynamics give rise to convex $ \calP_{i} $ and $ \widehat{\calX}_{i}^{f} $. However, even if dynamics~\eqref{eq:subsystem dynamics} are nonlinear, the local optimization problems~\eqref{eq:opt criterion}-\eqref{eq:opt constraints} approximately constitute a convex problem for sufficiently small~$ \widehat{\calC}_{i} $. This is because consistency constraint~\eqref{seq:opt constraints 2} restricts the optimal solutions to a neighborhood of a reference trajectory in which the system dynamics are approximately linear.

Moreover, we do not require that reference trajectories $ x_{i\in\calV}^{\text{ref}}[\cdot|k] $ satisfy any dynamics (cf. Ass.~\ref{ass:consistency constraint}) after initialization, i.e., for $ k>0 $. This is in contrast to other DMPC approaches employing consistency constraints~\cite{Dunbar2007,farina2014,farina2012}. Moreover, we allow that reference trajectories can be updated at every time step which allows for enhanced performance compared to algorithms with fixed reference trajectories (see performance comparison Sec.~\ref{subsec:sim performance analysis}). We detail the reference trajectory update in the next section. 

\section{Parallelized DMPC Algorithm}
\label{sec:main results - algorithm}

In the previous section, we have formulated local optimization problems for all subsystems and assumptions that allow for their parallelized evaluation while ensuring recursive feasibility. While the satisfaction of Ass.~\ref{ass:common coupled state constraints},~\ref{ass:existence of RPI set},~\ref{ass:terminal set} needs to be ensured during the initialization, Ass.~\ref{ass:consistency constraint} is the only assumption that needs to be taken into account online. In this section, we detail the initialization procedure, present a recursive algorithm to update reference trajectories such that Ass.~\ref{ass:consistency constraint} is satisfied, and state the overall DMPC algorithm.

\subsection{Initialization}
\label{subsec:init}

In order to solve the local optimization problems, some parameters need to be chosen offline. Therefore, we suggest the following initialization procedure: 

\emph{Step~1 (Problem Formulation):} Formulate the control problem such that neighboring agents have those constraints that couple their states in common (Ass.~\ref{ass:common coupled state constraints}). This can be always achieved by adding coupled state constraints to neighboring subsystems.

\emph{Step~2 (RPI sets):} For each subsystem $ i\in\calV $, determine an RPI set $ \calP_{i} $ and the corresponding state-feedback controller $ K_{i} $. For each subsystem, the computations are independent of the other subsystems due to the decoupled dynamics. Available methods are reviewed in Remark~\ref{remark:rpi sets}. The existence of $ \calP_{i} $ and $ K_{i} $, $ i\in\calV $, is assumed in Ass.~\ref{ass:existence of RPI set}.

\emph{Step~3 (Terminal constraints, stage cost function, and consistency constraint set):} For each subsystem $ i\in\calV $, choose symmetric positive-definite matrices $ \bm{Q}_{i}, \bm{R}_{i} $ and a stage cost function $ l_{i}(x,u) = ||x\!-\!\xi_{i}||^{2}_{\bm{Q}_{i}}\!+\!||u\!-\!u_{\xi_{i}}||^{2}_{\bm{R}_{i}} $. Thereafter, determine terminal sets $ \widehat{\calX}_{i}^{f} $ and corresponding terminal cost functions $ J_{i}^{f} $ such that Ass.~\ref{ass:terminal set} holds. 
To this end, we first construct auxilliary sets $ \widetilde{\calX}_{i}^{f} $, $ i\in\calV $, that satisfy Ass.~\ref{ass:terminal set a2}-\ref{ass:terminal set a4}. These can be computed independently for each subsystem due to the decoupled dynamics. For systems whose linearization is stabilizable, an approach using the discrete-time algebraic Riccati equation can be chosen. Let 
\begin{align*}
	\hat{f}_{i}(\hat{x}_{i},\hat{u}_{i}) = \bm{A}_{i} \hat{x}_{i} + \bm{B}_{i} \hat{u}_{i} + \hat{f}'_{i}(\hat{x}_{i},\hat{u}_{i})
\end{align*}
for some matrices $ \bm{A}_{i} $, $ \bm{B}_{i} $ with respective dimensions and $ \hat{f}'_{i}: \bbR^{n_{i}} \times \bbR^{m_{i}} \rightarrow \bbR^{n_{i}} $. By solving the discrete-time algebraic Riccati equation 
\begin{align*}
	\bm{P}_{i} = &\bm{A}_{i}^{T}\bm{P}_{i}\bm{A}_{i}\\&-(\bm{A}_{i}^{T}\bm{P}_{i}\bm{B}_{i})(\bm{R}_{i}+\bm{B}_{i}^{T}\bm{P}_{i}\bm{B}_{i})^{-1}(\bm{B}_{i}^{T}\bm{P}_{i}\bm{A}_{i})+\bm{Q}_{i},
\end{align*}
a positive definite matrix $ \bm{P}_{i} $ is obtained. Choose $ k_{i}^{f}(x_{i}) = -(\bm{R}_{i}+\bm{B}_{i}^{T}\bm{P}_{i}\bm{B}_{i})^{-1}\bm{B}_{i}^{T}\bm{P}_{i}\bm{A}_{i}x_{i} $ and $ J_{i}^{f}(x_{i}) = \sigma_{i} x_{i}^{T}\bm{P}_{i}x_{i}  $ with a scalar $ \sigma_{i} \geq 0 $. Then, there exists a sufficiently small scalar $ \tilde{\gamma}_{i}\geq0 $ such that Ass.~\ref{ass:terminal set a2},~\ref{ass:terminal set a4} hold for all $ x_{i}\in \widetilde{\calX}_{i}^{f}=\lbrace x | J_{i}^{f}(x) \leq \tilde{\gamma}_{i}\rbrace $ \cite[Remark~5.15]{gruene2011}. At last, choose $ \widehat{\calC}_{i} $ as some neighborhood of the origin and set $ \calC_{i} = \widehat{\calC}_{i} \oplus \calP_{i} $ (this choice might need to be refined later). Then select in a centralized way\footnote{Also a distributed computation of $ \gamma_{i\in\calV} $ is possible. For example, each subsystem incrementally increases $ \gamma_{i} $ and checks at each step constraint satisfaction. For each subsystem, the largest $ \gamma_{i} $ is chosen that still satisfies the constraints. Alternatively, distributed iterative optimization algorithms can be considered \cite{Falsone2017}.} $ \widehat{\calX}_{i}^{f}=\lbrace x | J_{i}^{f}(x) \leq {\gamma}_{i}\rbrace \subseteq \widetilde{\calX}_{i}^{f} $ with a sufficiently small scalar $ \gamma_{i}\in[0,\tilde{\gamma}_{i}] $ such that the constraints in Ass.~\ref{ass:terminal set a1} hold for all $ i\in\widehat{\calX}^{f}_{i} $ and all subsystems $ i\in\calV $. Note that Ass.~\ref{ass:terminal set a3} is satisfied due to the choice of $ \widehat{\calX}_{i}^{f} $ as a super-level set of~$ J_{i}^{f} $.

\emph{Step~4 (Initially feasible reference trajectories):} Determine initially feasible reference trajectories $ x_{i}^{\text{ref}}[\cdot|0] $, $ i\in\calV $, that satisfy~\eqref{eq:opt init trajectories}, e.g. by solving a centralized optimization problem subject to~\eqref{eq:opt init trajectories}. If no such trajectories could be found, choose a smaller set $ \widehat{\calC}_{i} $. 
\\
Note that \eqref{seq:opt init trajectories 2}-\eqref{seq:opt init trajectories 3} can often be simplified. If $ h_{i}, c_{i,r} $ are \emph{linear} and $ \calC_{i\in\calV} $ are polytopes, then \eqref{seq:opt init trajectories 2}-\eqref{seq:opt init trajectories 3} can be expressed as linear algebraic inequalities. Libraries for computations with polytopes are MPT3 (Matlab) \cite{Herceg2013}, Polyhedra (Julia)~\cite{Legat2021} and Polytope (Python) \cite{Polytope}. Alternatively, if $ h_{i}, c_{i,r} $ are nonlinear but still \emph{Lipschitz continuous}, then there exist scalars $ H_{i} $ and $ C_{i,r} $ such that\footnote{For simplicity, $ h_{i} $ and $ c_{i,r} $ are assumed to be scalar. In the case of vectors, subsequent calculations hold for row-wise evaluation.}
\begin{align*}
	|h_{i}(x')-h_{i}(x'')| &\leq H_{i} ||x'-x''|| \\
	|c_{i,r}(y')-c_{i,r}(y'')| &\leq C_{i,r}||y'-y''||
\end{align*}
where $ x',x'' \in \bbR^{n_{i}} $ and $ y', y'' \in \bigtimes_{j\in\lbrace i\rbrace\cup\calN_{i,r_{i}}} \bbR^{n_{j}} $. Define the maximal distance of any point in some compact set $ \calA\subset\bbR^{n} $ to some point $ x\in\bbR^{n} $ as $ {d}_{\text{max}}(x,\calA) := \sup_{z\in\calA} ||x-z||. $
Then, there exist scalars $ \nu_{h_{i}} := {d}_{\text{max}}(\mathbf{0},\calC_{i}) $ and $ \nu_{c_{i,r}} := {d}_{\text{max}}(\mathbf{0},\bigtimes_{j\in\lbrace i\rbrace\cup\calN_{i,r}}\calC_{j}) $, and it holds
\begin{align*}
	&\hspace{0.2cm}\sup_{x\in x_{i}^{\text{ref}}\oplus\calC_{i}}\hspace{-0.0cm} |h_{i}(x_{i}^{\text{ref}}[\kappa|0])-h_{i}(x)| \leq H_{i}\nu_{h_{i}} \\
	&\hspace{-0.2cm}\sup_{\hspace{0.2cm}x\in \hspace{-0.3cm}\bigtimes\limits_{\tiny\substack{\hspace{0.1cm}j\in\lbrace i \rbrace\cup\calN_{i,r}}}\hspace{-0.5cm} x^{\text{ref}}_{j}\oplus\calC_{j}} \hspace{-0.3cm}|c_{i,r}(x^{\text{ref}}_{j\in\lbrace i\rbrace\!\cup\!\calN_{i,r}}\![\kappa|0])\!-\!c_{i,r}(x)| \!\leq\! C_{i,r}\nu_{c_{i,r}}.
\end{align*}
Thus, we can replace \eqref{seq:opt init trajectories 2}-\eqref{seq:opt init trajectories 3} by
\begin{align*}
	h_i(x^{\text{ref}}_i[\kappa|0]) &\!\leq\! -H_{i}\nu_{h_{i}} \\
	c_{i,r}(x^{\text{ref}}_i[\kappa|0],x^{\text{ref}}_{j\in\calN_{i,r}}\![\kappa|0]) &\!\leq\! -C_{i,r}\nu_{c_{i,r}}.
\end{align*}

\subsection{Online Determination of Reference Trajectories}
\label{subsec:detremination of reference states}

In Ass.~\ref{ass:consistency constraint}, general conditions are stated that allow for the parallelized evaluation of the local optimization problems \eqref{eq:opt criterion} subject to~\eqref{eq:opt constraints} while preserving recursive feasibility (cf. Thm.~\ref{theorem:main theorem}) and ensuring state constraint satisfaction (cf. Prop.~\ref{lemma:constraint satisfaction}). Algorithms that ensure the satisfaction of Ass.~\ref{ass:consistency constraint} are essential to the proposed DMPC scheme. The generality of Ass.~\ref{ass:consistency constraint} allows for various such algorithms. In this section, we propose such an algorithm that updates a previous reference trajectory for each subsystem $ i\in\calV $. The algorithm is distributed. 

In particular, let each subsystem $ i\in\calV $ be initialized with a set of initially feasible reference trajectories $ x_{i}^{\text{ref}}[\cdot|0] $. Then for $ k\geq0 $, each subsystem $ i\in\calV $ checks for all $ \kappa\in\lbrace k+1,\dots,k+N \rbrace $ separately if 
\begin{align}
	\label{eq:determination ref states state constraints}
	h_i(x_i) \leq \mathbf{0} \qquad \forall x_i \in \hat{x}_i^{\ast}[\kappa|k]\oplus\calC_i
\end{align}
and
\begin{align}
	\label{eq:determination ref states coupled state constraints}
	\begin{split}
		&c_{i,r}(x_i,x_{j\in\calN_{i,r}}) \leq \mathbf{0} \qquad \forall x_i \in \hat{x}_i^{\ast}[\kappa|k]\oplus\calC_i, \\
		&\;\forall x_{j} \in (\hat{x}_j^{\ast}[\kappa|k]\oplus\calC_j) \cup (x_j^{\text{ref}}[\kappa|k]\oplus\calC_j), \; j\in\calN_{i,r}
	\end{split}
\end{align}
for all $ r=1,\dots,R_{i} $. 

\begin{rem}
	From a practical point of view, \eqref{eq:determination ref states state constraints}-\eqref{eq:determination ref states coupled state constraints} can be efficiently evaluated analogously to Step~4 in the previous section. Following the same reasoning, if $ h_{i} $, $ c_{i,r} $ are \emph{linear} and $ \calC_{i\in\calV} $ are polytopes, then \eqref{eq:determination ref states state constraints}-\eqref{eq:determination ref states coupled state constraints} can be expressed as linear algebraic inequalities. Alternatively, if $ h_{i}, c_{i,r} $ are only \emph{Lipschitz continuous}, then \eqref{eq:determination ref states state constraints}-\eqref{eq:determination ref states coupled state constraints} can be replaced by
	\begin{align*}
		h_{i}(\hat{x}_{i}^{\ast}[\kappa|k]) &\leq -H_{i}\nu_{h_{i}} 	\\
		c_{i,r}(\hat{x}_{i}^{\ast}[\kappa|k],\hat{x}_{j\in\calN_{i,r}}^{\ast}[\kappa|k]) &\leq -C_{i,r} \nu_{c_{i,r}} \\ 	c_{i,r}(\hat{x}_{i}^{\ast}[\kappa|k],x_{j\in\calN_{i,r}}^{\text{ref}}[\kappa|k]) &\leq -C_{i,r} \nu_{c_{i,r}} 
	\end{align*}
	where $ H_{i} $, $ C_{i,r} $, $ \nu_{h_{i}} $ and $ \nu_{c_{i,r}} $ are as before. In other cases, the evaluation is more elaborate.
\end{rem}

Based on conditions \eqref{eq:determination ref states state constraints}-\eqref{eq:determination ref states coupled state constraints}, the updated reference trajectories are defined for each subsystem $ i\in\calV $ and $ \kappa = k\!+\!1,\dots,k\!+\!N\!-\!1 $ as
\begin{subequations}
	\label{eq:reference update}
	\begin{align}
		\label{seq:reference update 1}
		x_{i}^{\text{ref}}[\kappa|k\!+\!1] &:= \!
		\begin{cases}
			\hat{x}_{i}^{\ast}[\kappa|k] & \text{if \eqref{eq:determination ref states state constraints}-\eqref{eq:determination ref states coupled state constraints} hold for $ \kappa $,} \\
			x_{i}^{\text{ref}}[\kappa|k] & \text{otherwise,}
		\end{cases} \\
		\label{seq:reference update 2}
		x_{i}^{\text{ref}}[k\!+\!N|k\!+\!1] &:= \hat{x}_{i}^{\ast}[k\!+\!N|k].
	\end{align}
\end{subequations}

Intuitively, each subsystem attempts to change its previous reference state $ x_{i}^{\text{ref}}[\kappa|k] $ to the predicted state $ \hat{x}_{i}^{\ast}[\kappa|k] $ for any $ \kappa $. The reference states, however, are only changed if no state within a $ \calC_{i} $-neighborhood of the predicted state violates any of the state constraints~\eqref{eq:state constraints} which is checked by conditions \eqref{eq:determination ref states state constraints} and \eqref{eq:determination ref states coupled state constraints}. In condition~\eqref{eq:determination ref states coupled state constraints} on the coupled state constraints, both the predicted states $ \hat{x}_{j\in\calN_{i,r}}^{\ast} $ and the previous reference states $ x_{j\in\calN_{i,r}}^{\text{ref}} $ of the neighboring subsystems are considered. This allows for the parallelized computation of reference trajectories. The reference trajectory update of subsystem~$ i $ for $ k\geq0 $ is summarized in Algorithm~\ref{algo:reference states}. It guarantees the satisfaction of Ass.~\ref{ass:consistency constraint} as stated by the following proposition.

\begin{algorithm}[t]
	\caption{Reference Trajectory Update of Subsystem $ i $}
	\textbf{Input} $ \hat{x}_{i}^{\ast}[\cdot|k] $, $ \hat{x}_{j\in\calN_{i}}^{\ast}[\cdot|k] $, $ x_{i}^{\text{ref}}[\cdot|k] $, $ x_{j\in\calN_{i}}^{\text{ref}}[\cdot|k] $ \\
	\textbf{Output} $ x_{i}^{\text{ref}}[\cdot|k+1] $
	\begin{algorithmic}[1]
		\For{$ \kappa= k+1,\dots,k+N-1$}
			\If{\eqref{eq:determination ref states state constraints} and \eqref{eq:determination ref states coupled state constraints} hold for all $ r=1,\dots,R_{i} $}
				\State $ x_{i}^{\text{ref}}[\kappa|k\!+\!1] := \hat{x}_{i}^{\ast}[\kappa|k]  $;
			\Else
				\State $ x_{i}^{\text{ref}}[\kappa|k\!+\!1] := x_{i}^{\text{ref}}[\kappa|k]  $;
			\EndIf
		\EndFor
		\State $ x_{i}^{\text{ref}}[k+N|k\!+\!1] := \hat{x}_{i}^{\ast}[k+N|k]  $;
	\end{algorithmic}
	\label{algo:reference states}
\end{algorithm}

\begin{prop}
	\label{prop:determination ref states}
	Let Ass.~\ref{ass:common coupled state constraints} and Ass.~\ref{ass:terminal set a1} be satisfied, and let there be a set of initially feasible reference trajectories $ x^{\text{ref}}_{i}[\cdot|0] $, $ i\in\calV $. Then for all $ i\in\calV $, reference trajectories $ x_{i}^{\text{ref}}[\cdot|0] $ and $ x_{i}^{\text{ref}}[\cdot|k] $, $ k>0 $, recursively defined in~\eqref{eq:reference update}, satisfy Ass.~\ref{ass:consistency constraint} for all times $ k\geq0 $.
\end{prop}
\begin{pf}
	\emph{Satisfaction of Ass.~\ref{ass:consistency constraint a1}:} At first, we show by induction that $ x_{i}^{\text{ref}}[\cdot|k] $ satisfies Ass.~\ref{ass:consistency constraint a1} for all $ k\geq0 $. Note that initially feasible reference trajectories $ x_{i}^{\text{ref}}[\cdot|0] $ satisfy Ass.~\ref{ass:consistency constraint a1} for $ k \!=\! 0 $ by definition.
	Next, assuming that $ x_{i}^{\text{ref}}[\kappa|k] $, $ \kappa=k,\dots,k\!+\!N\!-\!1 $, satisfies Ass.~\ref{ass:consistency constraint a1}, we show that also $ x_{i}^{\text{ref}}[\kappa|k\!+\!1] $, $ \kappa=k\!+\!1,\dots,k\!+\!N $, satisfies Ass.~\ref{ass:consistency constraint a1}. We split the proof in two parts: at first, we show that $ x_{i}^{\text{ref}}[\kappa|k\!+\!1] $, $ \kappa=k\!+\!1,\dots,k\!+\!N\!-\!1 $, as defined in~\eqref{seq:reference update 1} satisfies Ass.~\ref{ass:consistency constraint a1}. Thereafter, we show that also $ x_{i}^{\text{ref}}[k\!+\!N|k\!+\!1] $ as defined in~\eqref{seq:reference update 2} satisfies Ass.~\ref{ass:consistency constraint a1} for $ \kappa = k\!+\!N $. From this, we can conclude, that $ x_{i}^{\text{ref}}[\cdot|k\!+\!1] $ overall satisfies Ass.~\ref{ass:consistency constraint a1} at time $ k\!+\!1 $.
	
	\emph{Part 1:} Consider $ x_{i}^{\text{ref}}[\kappa|k\!+\!1] $, $ \kappa=k\!+\!1,\dots,k\!+\!N\!-\!1 $, as defined in~\eqref{seq:reference update 1}. Note that $ {x}_i^{\text{ref}}[\kappa|k\!+\!1]\in\lbrace \hat{x}_i^{\ast}[\kappa|k], x_i^{\text{ref}}[\kappa|k] \rbrace $ for all $ i\in\calV $. We consider two cases:\\	
	\emph{Case 1:} Let conditions~\eqref{eq:determination ref states state constraints}-\eqref{eq:determination ref states coupled state constraints} be satisfied. Then observe that \eqref{eq:determination ref states state constraints} is equivalent to \eqref{eq:ref trajectory state constraint}. Moreover, as for any neighbor $ j\in\calN_{i} $, it holds that $ x_j^{\text{ref}}[\kappa|k+1]\!\in\!\lbrace \hat{x}_j^{\ast}[\kappa|k], {x}_j^{\text{ref}}[\kappa|k] \rbrace $, \eqref{eq:determination ref states coupled state constraints} implies~\eqref{eq:ref trajectory coupled state constraint}. We conclude the satisfaction of Ass.~\ref{ass:consistency constraint a1} in case~1. \\		
	 \emph{Case 2:} If conditions~\eqref{eq:determination ref states state constraints}-\eqref{eq:determination ref states coupled state constraints} are \emph{not} satisfied for~$ \kappa $, then $ x_i^{\text{ref}}[\kappa|k\!+\!1] = x_i^{\text{ref}}[\kappa|k] $. As $ x_i^{\text{ref}}[\kappa|k] $ satisfies \eqref{eq:ref trajectory state constraint} at time-step $ k $, it does so at~$ k\!+\!1 $ since \eqref{eq:ref trajectory state constraint} is time-invariant and does not depend on other subsystems. 
	 In order to determine the satisfaction of~\eqref{eq:ref trajectory coupled state constraint}, we assume at first that conditions~\eqref{eq:determination ref states state constraints}-\eqref{eq:determination ref states coupled state constraints} are \emph{not} satisfied at $ \kappa $ and any of the neighbors $ j\in\calN_{i} $. Then $ x_j^{\text{ref}}[\kappa|k\!+\!1] = x_j^{\text{ref}}[\kappa|k] $ for all $ j\in\calN_{i} $, and the  satisfaction of~\eqref{eq:ref trajectory coupled state constraint} trivially follows from the previous time step~$ k $ when $ x_j^{\text{ref}}[\kappa|k] $ satisfies \eqref{eq:ref trajectory coupled state constraint} by assumption. Next, assume that for all $ j\in\widetilde{\calN}_{i} $, where $ \widetilde{\calN}_{i}\subseteq\calN_{i} $ is an arbitrary subset of $ \calN_{i} $, conditions~\eqref{eq:determination ref states state constraints}-\eqref{eq:determination ref states coupled state constraints} are satisfied. Then, $ x_j^{\text{ref}}[\kappa|k\!+\!1] = \hat{x}_j^{\ast}[\kappa|k] $ for all $ j\in\widetilde{\calN}_{i} $. Let $ r $ be any $ r\in\lbrace 1,\dots,R_i\rbrace $ and consider the coupled state constraint $ c_{i,r} $. By Ass.~\ref{ass:common coupled state constraints}, there exist constraints $ c_{j,r'} $, $ r'\in\lbrace 1,\dots,R_{j} \rbrace $, for all neighbors $ j\in\calN_{i,r} $ of subsystem~$ i $ such that 
	\begin{align}
		\label{eq:prop ref update eq1}
		c_{j,r'}(x_j, x_{j'\in\calN_{j,r'}}) \equiv c_{i,r}(x_i,x_{i'\in\calN_{i,r}}).
	\end{align}
	For these $ c_{j,r'} $, the satisfaction of~\eqref{eq:determination ref states coupled state constraints} implies 
	\begin{align*}
		c_{j,r'}(x_j,x_{{j'}\in\calN_{j,r'}}) \leq \mathbf{0} \qquad \forall x_j \in x_j^{\text{ref}}[\kappa|k\!+\!1]\oplus\calC_j,& \\
		\hspace{2cm}\forall x_{j'} \in x_{{j'}}^{\text{ref}}[\kappa|k\!+\!1]\oplus\calC_{{j'}}, \; {j'}\in\calN_{j,r'}.&
	\end{align*}
	Due to \eqref{eq:prop ref update eq1} and since $ i\in\calN_{j,r'} $, we can rewrite the latter equation as
	\begin{align*}
		c_{i,r}(x_i,x_{{j}\in\calN_{i,r}}) \leq \mathbf{0} \qquad \forall x_i \in x_i^{\text{ref}}[\kappa|k\!+\!1]\oplus\calC_i,& \\
		\hspace{2cm}\forall x_j \in x_{{j}}^{\text{ref}}[\kappa|k\!+\!1]\oplus\calC_{{j}}, \; {j}\in\calN_{i,r}&
	\end{align*} 
	where $ \calN_{i,r} = (\calN_{j,r'}\!\cup\!\lbrace j \rbrace)\!\setminus\lbrace i \rbrace $. This is equal to \eqref{eq:ref trajectory coupled state constraint} at time $ k\!+\!1 $. Thus, we have shown that even in the case that conditions~\eqref{eq:determination ref states state constraints}-\eqref{eq:determination ref states coupled state constraints} are satisfied for some of the neighbors of subsystem~$ i $, namely $ j\in\widetilde{\calN}_{i} $, \eqref{eq:ref trajectory coupled state constraint} still holds. Thereby, Ass.~\ref{ass:consistency constraint a1} is also satisfied in case~2.
	
	\emph{Part 2:} Consider $ x_{i}^{\text{ref}}[\kappa|k+1] $ at $ \kappa\!=\!k\!+\!N $ where $ x_{i}^{\text{ref}}[k\!+\!N|k\!+\!1] := \hat{x}_{i}^{\ast}[k\!+\!N|k] $ by~\eqref{seq:reference update 2}. Since $ \hat{x}_i^{\ast}[k\!+\!N|k]\in\widehat{\calX}_i^{f} $ according to terminal constraint \eqref{seq:opt constraints 4}, the choice $ x_i^{\text{ref}}[\kappa|k\!+\!1] $ satisfies Ass.~\ref{ass:consistency constraint a1} due to Ass.~\ref{ass:terminal set a1}. 
	
	\emph{Satisfaction of Ass.~\ref{ass:consistency constraint a2}:} At last, we show the satisfaction of Ass.~\ref{ass:consistency constraint a2}. To this end, we consider the two cases for $ x_{i}^{\text{ref}}[\kappa|k\!+\!1] $, $ \kappa=k\!+\!1,\dots,k\!+\!N\!-\!1 $, from part~1 again. In case~1, if conditions~\eqref{eq:determination ref states state constraints}-\eqref{eq:determination ref states coupled state constraints} are satisfied  for $ \kappa $, then $ x_i^{\text{ref}}[\kappa|k\!+\!1] = \hat{x}_i^{\ast}[\kappa|k] $ which trivially implies \eqref{eq:ref trajectory recursive feasibillity constraint}. In case~2, if conditions~\eqref{eq:determination ref states state constraints}-\eqref{eq:determination ref states coupled state constraints} are \emph{not} satisfied, we have 
	\begin{align*}
		\hat{x}^{\ast}_i[\kappa | k] \stackrel{\eqref{seq:opt constraints 2}}{\in} x^{\text{ref}}_i[\kappa|k ]\oplus\calC_i = x^{\text{ref}}_i[\kappa|k\!+\!1 ]\oplus\calC_i
	\end{align*} 
	which is equivalent to~\eqref{eq:ref trajectory recursive feasibillity constraint}. For $ x_{i}^{\text{ref}}[\kappa|k+1] $ at $ \kappa\!=\!k\!+\!N $, \eqref{eq:ref trajectory recursive feasibillity constraint} is trivially satisfied. Altogether, we also conclude the satisfaction of Ass.~\ref{ass:consistency constraint a2}. 
	{\color{white}.}\hfill$ \Box $
\end{pf}


Other strategies to update the reference trajectories are also possible; e.g., sets $ \calC_{i} $ could be varied in addition.


\subsection{Distributed MPC Algorithm}
\label{subsec:Distributed MPC Algorithm}

The distributed MPC algorithm is initialized as detailed in Sec.~\ref{subsec:init}. Then, the distributed MPC algorithms as given in Algorithm~\ref{algo:DMPC} are executed in parallel by all subsystems $ i\in\calV $. Note that Algorithm~\ref{algo:reference states} can be replaced by any other algorithm that ensures the satisfaction of Ass.~\ref{ass:consistency constraint} for all $ k\geq0 $. 

\begin{algorithm}[t]
	\caption{Distributed MPC Algorithm for Subsystem $ i $}
	\textbf{Input} 	$ \xi_{i}, \calP_{i}, \calC_{i}, K_i, l_i, J_{i}^{f}, \widehat{\calX}_{i}^{f}, x_{i}^{\text{ref}}[\cdot|0] $
	\begin{algorithmic}[1]
		\State $ k \gets 0 $;
		\While{$ k\geq 0 $}
			\State measure $ x_{i}[k] $;
			\State solve~\eqref{eq:opt criterion} subject to~\eqref{eq:opt constraints};
			\State communicate $ \hat{x}_{i}^{\ast}[\cdot|k] $ to all neighbors $ j\in\calN_{i} $;
			\State receive $ \hat{x}_{j}^{\ast}[\cdot|k] $, $ j\in\calN_i $;
			\State apply $ u_i^{\text{MPC}} $ as given in~\eqref{eq:actual input law};
			\State compute $ x_{i}^{\text{ref}}[\cdot|k+1] $ with Algorithm~\ref{algo:reference states};
			\State communicate $ \hat{x}_{i}^{\text{ref}}[\cdot|k+1] $ to all neighbors $ j\in\calN_{i} $;
			\State receive $ \hat{x}_{j}^{\text{ref}}[\cdot|k+1] $, $ j\in\calN_i $;
			\State $ k\gets k+1 $;
		\EndWhile
	\end{algorithmic}
	\label{algo:DMPC}
\end{algorithm}

\begin{rem}[Iterative version]
	\label{remark:iterative dmpc}
	If Algorithm~\ref{algo:DMPC} is initialized with suboptimal initially feasible reference trajectories, then an iterative version of the proposed DMPC scheme can lead to an improved performance. Therefore, Algorithm~\ref{algo:DMPC} is modified as follows: After solving local optimization problem~\eqref{eq:opt criterion}-\eqref{eq:opt constraints}, a new reference trajectory $ x_{i}^{\text{ref}, +}[\kappa|k] $,  $ \kappa\!=\!k,\dots,k\!+\!N\!-\!1 $, is computed as
	\begin{align*}
		x_{i}^{\text{ref}, +}[\kappa|k] &:= \!
		\begin{cases}
			\hat{x}_{i}^{\ast}[\kappa|k] & \text{\eqref{eq:determination ref states state constraints}-\eqref{eq:determination ref states coupled state constraints} hold for $ \kappa $,} \\
			x_{i}^{\text{ref}}[\kappa|k] & \text{otherwise.}
		\end{cases}
	\end{align*}
	Then, the local optimization problems are repeatedly solved. Depending on the available computation time, this procedure can be repeated multiple times, before applying the last computed control input $ u_{i}^{\text{MPC}}[k] $. An example is presented in Sec.~\ref{sec:simulation}.
\end{rem}

\begin{rem}[Dynamic couplings]
	The proposed approach is not directly generalizable to systems with dynamic couplings. The main challenges are twofold. Firstly, a parallelized non-iterative DMPC scheme presumably only yields approximate asymptotic convergence to a neighborhood of the desired states, which~\cite{Dunbar2007} alleges. Secondly, the heterogeneity in the subsystems' constraints hardens the update of the reference trajectories (to circumvent this, \cite{farina2012,farina2014} employ fixed reference trajectories). It is expected that further conditions that account for this heterogeneity in the case of systems with bounded dynamic couplings need to be invoked to ensure recursive feasability. This however is a non-trivial problem on its own and left for future research. For systems with bounded dynamic couplings, however, the proposed DMPC scheme can be extended (though conservatively) along the lines of \cite[Sec.~4.6]{Wiltz2023b}.
\end{rem}



\section{Simulation}
\label{sec:simulation}

In this section, we investigate the performance of the proposed algorithm with respect to computation time and optimality. To this end, we consider mobile robots subject to connectivity and collision avoidance constraints, which are often considered coupled state constraints in the literature \cite{Bono2021,Hagen2018,Nikou2019}. 

In particular, we consider the kinematic model of three-wheeled omni-directional robots. The state of robot~$ i $ is given as $ \mathbf{x}_i \!:=\! [x_{i}, y_{i},\psi_i] $ where $ x_{i} $, $ y_{i} $ denote the position coordinates and $ \psi_i $ the orientation; its position is defined as $ \mathbf{x}^{\text{pos}}_{i} \!:=\! [x_{i},y_{i}]^T $. The dynamics of robot~$ i $ are given as
\begin{align}
	\label{eq:dynamics three wheeled robot}
	\dot{\mathbf{x}}_i = \bm{R}(\psi_i) \, (\bm{B}_i^{T})^{-1} \, r_i \, u_i + w_{i}
\end{align}
where 
\begin{align*}
	\bm{R}(\psi_i) \!=\! \bigg[
	\begin{smallmatrix}
	\cos(\psi_i) & -\sin(\psi_i) & 0 \\
	\sin(\psi_i) & \cos(\psi_i) & 0 \\
	0 & 0 & 1
	\end{smallmatrix} \bigg], \;
	\bm{B}_i \!= \!\bigg[
	\begin{smallmatrix}
	0 & \cos(\pi/6) & -\cos(\pi/6) \\
	-1 & \sin(\pi/6) & \sin(\pi/6) \\
	l_i & l_i & l_i
	\end{smallmatrix} \bigg],
\end{align*}
$ l_i = 0.2 $ is the radius of the robot body, $ r_i = 0.02 $ the wheel radius, $ {u}_i = [u_{i,1},u_{i,2},u_{i,3}]^T $ the angular velocity of the wheels, and $ w_{i} = [w_{i,1},w_{i,2},w_{i,3}]^T \in \calW_{i} \subset \bbR^{3} $ a bounded uniformly distributed disturbance. The corresponding nominal dynamics are 
\begin{align}
	\label{eq:nominal dynamics three wheeled robot}
	\dot{\hat{\mathbf{x}}}_i = \bm{R}(\hat{\psi}_i) \, (\bm{B}_i^{T})^{-1} \, r_i \, \hat{u}_i
\end{align}
where $ \hat{\mathbf{x}}_{i} \!:=\! [\hat{x}_{i}, \hat{y}_{i},\hat{\psi}_i] $ is the nominal state of robot~$ i $ and $ \hat{u}_i $ the nominal input. The nominal position is defined as $ \hat{\mathbf{x}}^{\text{pos}}_{i} \!:=\! [\hat{x}_{i},\hat{y}_{i}]^T $.

\subsection{Controller Design}

We follow the four initialization steps in Sec.~\ref{subsec:init} as presented next in great detail.
Let us consider three robots $ \calV = \lbrace 1,2,3 \rbrace $ with nonlinear dynamics \eqref{eq:dynamics three wheeled robot} which shall move from an initial formation~$ x_0 $ to a target formation~$ \xi $. 
All robots $ i\in\calV $ are subject to connectivity constraints
\begin{align}
	\label{eq:connectivity constraint}
	||\mathbf{x}^{\text{pos}}_{i} - \mathbf{x}^{\text{pos}}_{j}|| \leq d^{\text{max}}, \qquad j\in\calN_i:=\calV\setminus\lbrace i \rbrace
\end{align}
with $ d^{\text{max}} = 2.9 $, and input constraints $  ||u_i||_{\infty} \leq 15 $ where $ ||\cdot||_{\infty} $ denotes the maximum norm. It can be easily verified that the coupled state constraints satisfy Ass.~\ref{ass:common coupled state constraints} by design (step~1).
The continuous-time nominal dynamics \eqref{eq:nominal dynamics three wheeled robot} are discretized with time-step $ \Delta t = 1/3 $ using a 4th-order Runge-Kutta integration algorithm where the nominal inputs are applied as zero-order hold. The control input applied to the robots is 
\begin{align}
	\label{eq:sim example actual input}
	u_{i}(t) = \hat{u}_{i}(t) + K_{i}(\mathbf{x}_{i}(t), \hat{\mathbf{x}}_{i}(t))
\end{align}
where $ \hat{u}_{i}(t) = \hat{u}_{i}^{\text{MPC}}(\mathbf{x}_{i}[k]) $, $ t\in[k\,\Delta t, (k\!+\! 1)\Delta t) $. 

Next, we construct RPI sets $ \calP_{i} $, $ i\in\calV $ (step~2). To this end, we choose the continuous-time auxiliary controller~$ K_{i} $ as
\begin{align}
	\label{eq:sim example auxiliary controller}
	\begin{split}
		K_{i}(\mathbf{x}_{i}(t), \hat{\mathbf{x}}_{i}(t)) &= \bm{B}_{i}^{T} \bm{R}(-\psi_{i})\, \bm{\Lambda}_{i}\, (\mathbf{x}_{i}(t) - \hat{\mathbf{x}}_{i}(t))\\
		&= \bm{B}_{i}^{T} \bm{R}(-\psi_{i}) \, \bm{\Lambda}_{i} \, p_{i}(t)
	\end{split}
\end{align}
where $ \bm{\Lambda}_{i}\in\bbR^{3\times 3} $ is Hurwitz. Substituting \eqref{eq:sim example auxiliary controller} into \eqref{eq:sim example actual input}, \eqref{eq:sim example actual input} into \eqref{eq:dynamics three wheeled robot}, and computing $ \dot{p}_{i} = \dot{\mathbf{x}}_{i} - \dot{\hat{\mathbf{x}}}_{i} $ yields
\begin{align}
	\label{eq:sim example deviation dynamics continuous}
	\dot{p}_{i}(t) = \bm{\Lambda}_{i} p_{i}(t) + w_{i}(t).
\end{align}
We choose $ \bm{\Lambda}_{i} = -\text{diag}(6,6,5.5) $ for all $ i\in\calV $. To numerically compute the RPI set $ \calP_{i} $, we exactly discretize the continuous-time deviation dynamics \eqref{eq:sim example deviation dynamics continuous} and obtain
\begin{align}
	\label{eq:sim example deviation dynamics discrete}
	{p}_{i}[k+1] = \bm{\Lambda}_{i}^{\text{d}} p_{i}[k] + w_{i}[k]
\end{align}
with $ \bm{\Lambda}_{i}^{\text{d}} = e^{\bm{\Lambda}_{i} \Delta t} $, and $ w_{i}[k] \in\calW_{i}^{d} := \int_{\tau=0}^{\Delta t} e^{\bm{\Lambda}_{i} \tau} d\tau \; \calW_{i} $. We assume that $ \calW_{i}^{d} = \lbrace w_{i} \, | \, |w_{i,1}| < 0.1, \; |w_{i,2}| < 0.1, \; |w_{i,3}| < \pi/32 \rbrace $, or equivalently, $ \calW_{i} = \lbrace w_{i} \, | \, |w_{i,1}| < 0.6940, \; |w_{i,2}| < 0.6940, \; |w_{i,3}| < 0.6429 \rbrace $. Then, we can compute $ \calP_{i} $ as in~\cite{Rakovic2004}. The resulting RPI set $ \calP_{i} $ is a box given as $ \calP_{i} = \lbrace p_{i} \, |\, |p_{i,1}| < \bar{p}_{i,1} = 0.1157, \; |p_{i,2}| < \bar{p}_{i,2} = 0.1157, |p_{i,3}| < \bar{p}_{i,3} = 0.1169 \rbrace $. An outer approximation of $ \Delta \calU_{i} $ is computed in accordance with~\eqref{eq:Delta U_i}. Then, Ass.~\ref{ass:existence of RPI set} is satisfied. For required set computations, we use the MPT3 toolbox \cite{Herceg2013} for Matlab and YALMIP~\cite{Lofberg2004}.  

Now, we choose the performance matrices as $ \bm{Q}_1 = \text{diag}(100,100,100) $, $ \bm{Q}_2 = \bm{Q}_3 = \text{diag}(1,1,50) $, $ \bm{R}_1 = \text{diag}(1,1,1) $, $ \bm{R}_2 = \bm{R}_3 = \text{diag}(5,5,5) $, and consistency constraint sets $ \widehat{\calC}_{i} = \lbrace \zeta\in\bbR^{3} \, | \, ||[\zeta_{1},\zeta_{2}]^{T}||\leq\sqrt{2}\,\bar{c}_{i} \rbrace $ for all $ i\in\calV $ where $ \bar{c}_{i}=0.125 $. Matrices $ \bm{Q}_{1} $ and $ \bm{R}_{1} $ are chosen such that subsystem~1 tends faster to its desired state than the other subsystems. Thereby, subsystem~1 attempts to violate the connectivity constraint, which is of interest in the performance analysis conducted later (Sec.~\ref{subsec:sim performance analysis}). The terminal cost functions $ J_{i\in\calV}^{f} $ and terminal sets $ \widehat{\calX}_{i\in\calV}^{f} $ are computed via the discrete-time algebraic Riccati equation as outlined in Sec.~\ref{subsec:init}, step~3, and Ass.~\ref{ass:terminal set a2}-\ref{ass:terminal set a4} are satisfied. By choosing terminal sets~$ \widehat{\calX}_{i\in\calV}^{f} $ sufficiently small, also Ass.~\ref{ass:terminal set a1} is satisfied.

At last, we determine initially feasible reference trajectories (step~4). Therefore, observe that $ c_{ij}({\mathbf{x}}_i,{\mathbf{x}}_j) = ||{\mathbf{x}}^{\text{pos}}_{i} - {\mathbf{x}}^{\text{pos}}_{j}|| - d^{\text{max}} $ is Lipschitz continuous with $ C=1 $ for all $ i,j\in\calV $. Then following the discussion on Lipschitz continuous constraints in Sec.~\ref{subsec:init}, step~4, we can rewrite~\eqref{seq:opt init trajectories 3} for all $ i,j\in\calV $ more conservatively as
\begin{align}
	\label{eq:simplified concistency constraint dist condition}
	c_{ij}(\hat{\mathbf{x}}_i,\hat{\mathbf{x}}_j) = ||\hat{\mathbf{x}}^{\text{pos}}_{i} - \hat{\mathbf{x}}^{\text{pos}}_{j}|| - d^{\text{max}} \leq - \nu_{c_{ij}}
\end{align}
where $ \nu_{c_{ij}} = 2(\sqrt{2}\,\bar{c}_i+\bar{p}_{i}^{\,\text{pos}}) $ with $ \bar{p}_{i}^{\,\text{pos}} :	= (\bar{p}_{i,1}^{2}\!+\!\bar{p}_{i,2}^{2})^{0.5} = 0.1636 $. By solving an optimization subject to~\eqref{eq:opt init trajectories}, we compute initially feasible reference trajectories $ x_{i}^{\text{ref}} $, $ i\in\calV $, where \eqref{seq:opt init trajectories 3} is implemented as~\eqref{eq:simplified concistency constraint dist condition}. 
To conclude the controller design, we implement the local optimization problems~\eqref{eq:opt criterion}-\eqref{eq:opt constraints} where consistency constraint~\eqref{seq:opt constraints 2} is implemented as a box constraint using the inner approximation $ \widebar{\calC}_{i} = \lbrace x \in\bbR^{3} \, | \, ||[x_1, x_2]||_{\infty} \leq \bar{c}_i \rbrace $ of $ \widehat{\calC}_{i} $. Reference trajectories are updated at every time-step by Algorithm~\ref{algo:reference states} which ensures the satisfaction of Ass.~\ref{ass:consistency constraint} (cf. Prop.~\ref{prop:determination ref states}). Thereby, the satisfaction of all assumptions is ensured by the controller design.

\subsection{Simulation Results}
The three mobile robots start in the initial formation $ \mathbf{x}_{0,1} = [-1,0,0]^T $, $ \mathbf{x}_{0,2} = [-3,1,7\pi/4]^T $, $ \mathbf{x}_{0,3} = [-3,-1,\pi/4 ]^T $, and move to the target formation $ \xi_1 = [\xi_{11},0,\pi]^T $, $ \xi_2 = [1,-1,\pi/4]^T $, $ \xi_3 = [1,1,7\pi/4]^T $ where $ \xi_{11}\in\lbrace 1.5,2.0,2.5,3.0 \rbrace $. Observe that for increasing $ \xi_{11} $, the inter-robot distances in the target formation increase. The prediction time is chosen as $ T=12s $ and the prediction horizon as $ N = 36 $. 

\begin{figure}[tp]
	\centering
	\subfigure[Trajectories of the disturbed system with $ w_{i} \in\calW_{i} $. The black dashed line denotes $ \hat{\mathbf{x}}_{i}{[}k|k{]} $, the colored dotted lines the actual trajectories.]{
		\def\svgwidth{0.75\columnwidth}
		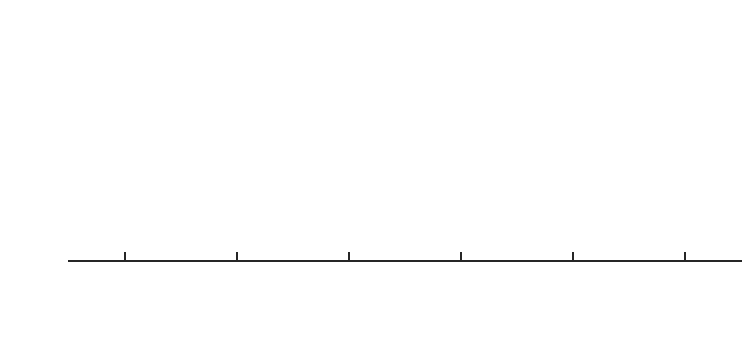
		\label{fig:trajectories}
	}
	\subfigure[Trajectories of the undisturbed system, i.e., $ w_{i} \equiv 0 $. The markers denote the robots' orientation. Nominal and actual trajectories coincide.]{
		\def\svgwidth{0.75\columnwidth}
		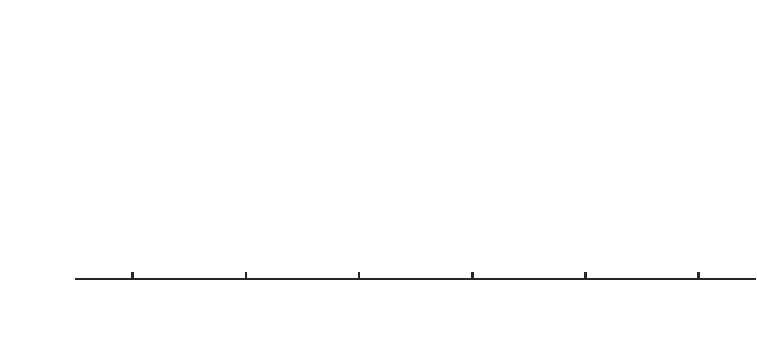
		\label{fig:trajectories_undisturbed_2p5}
	}
	\vspace{-0.3cm}
	\caption{Trajectories of robots for $ \xi_{11}=2.5 $.  }
	\label{fig:trajectories_both}
\end{figure}

\begin{figure}[tp]
	\centering
	\def\svgwidth{0.99\columnwidth}
	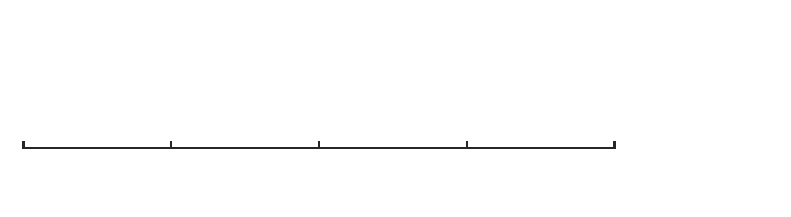
	\vspace{-0.3cm}
	\caption{Actual inter-robot distance $ d_{ij} = ||\hat{\mathbf{x}}^{\text{pos}}_{i} - \hat{\mathbf{x}}^{\text{pos}}_{j}|| $ for $ \xi_{11} = 3.0 $. The black line denotes $ d^{\text{max}} = 2.9 $.}
	\label{fig:actual_distance}
\end{figure}

\begin{figure*}[tbp]
	\centering
	\subfigure[$ \xi_{11} = 1.5 $]{
		\def\svgwidth{0.22\textwidth}
		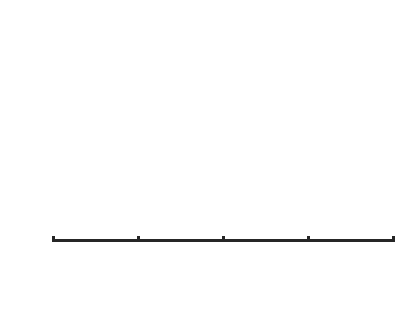
		\label{fig:distances_1p5}
	}
	\subfigure[$ \xi_{11} = 2.0 $]{
		\def\svgwidth{0.22\textwidth}
		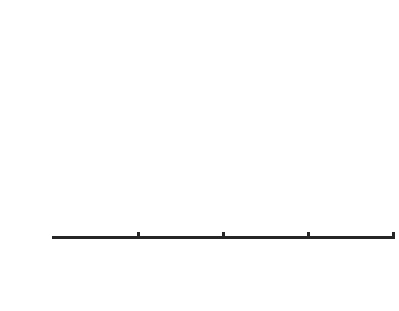
		\label{fig:distances_2p0}
	}
	\subfigure[$ \xi_{11} = 2.5 $]{
		\def\svgwidth{0.22\textwidth}
		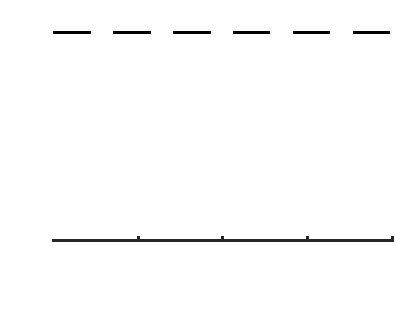
		\label{fig:distances_2p5}
	}
	\subfigure[$ \xi_{11} = 3.0 $]{
		\def\svgwidth{0.22\textwidth}
		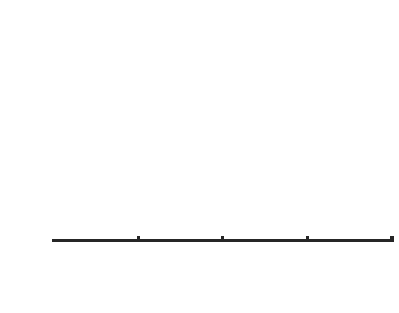
		\label{fig:distances_3p0}
	}
	\vspace{-0.3cm}
	\caption{Nominal inter-robot distances. Distance $ \hat{d}_{ij}{[}k|k{]} = ||\hat{\mathbf{x}}^{\text{pos}}_{i}{[}k|k{]} - \hat{\mathbf{x}}^{\text{pos}}_{j}{[}k|k{]}|| $ denotes the nominal distance between robot $ i $ and $ j $; the dashed line indicates $ d^{\text{max}} - \nu_{c_{ij}} $.}
	\label{fig:nominal_distances}
\end{figure*}

For $ \xi_{11} = 2.5 $, the resulting trajectories are depicted in Fig.~\ref{fig:trajectories_both}. Fig.~\ref{fig:trajectories} shows how the actual state trajectories $ \mathbf{x}_{i}[k] $ oscillate around the nominal trajectories $ \hat{\mathbf{x}}_{i}{[}k|k{]} $ due to the disturbances. For comparison, Fig.~\ref{fig:trajectories_undisturbed_2p5} shows the state trajectories in the absence of disturbances; for a detailed discussion of this case, we refer to~\cite{Wiltz2022c}. Fig.~\ref{fig:actual_distance} shows that the actual inter-robot distances satisfy the coupled state constraint~\eqref{eq:connectivity constraint}. 

\subsection{Performance Analysis}
\label{subsec:sim performance analysis}
In order to evaluate the performance of the proposed algorithm with respect to computation time and actual cost, we compare it with two other robust DMPC algorithms: (1)~Algorithm~\ref{algo:DMPC} with fixed reference trajectories
\begin{align*}
	x_{i}^{\text{ref}}[\kappa | k+1] := \hat{x}_{i}[\kappa | k]  \quad \text{for } \kappa=k+1,\dots,k+N.
\end{align*}
This choice of reference trajectories corresponds to the choice in~\cite{farina2012}. (2) Sequential DMPC~\cite{Nikou2019} which is based on \cite[Sec.~2]{Mueller2012} and~\cite{Richards2007}. The DMPC controllers are implemented using Casadi \cite{Andersson2019}, Ipopt and Matlab; simulations are performed on an Intel Core i5-10310U, 16GB~RAM. 

\begin{table}[tp]
	\centering
	\begin{tabularx}{0.5\textwidth}{|>{\centering\arraybackslash\hsize=0.4\hsize\linewidth=\hsize}X||>{\centering\arraybackslash\hsize=1.2\hsize\linewidth=\hsize}X|>{\centering\arraybackslash\hsize=1.2\hsize\linewidth=\hsize}X|>{\centering\arraybackslash\hsize=1.2\hsize\linewidth=\hsize}X|}
		\hline 
		$ \xi_{11} $ & Proposed DMPC & DMPC with fixed reference & Sequential DMPC \cite{Nikou2019}   \\ 
		\hline
		1.5 & 1.00,\,1.00,\,1.00 & 0.93,\,1.45,\,1.45 & 1.91,\,0.93,\,0.94\\
		\hline
		2.0 & 1.00,\,1.00,\,1.00 & 1.02,\,1.41,\,1.37 & 1.75,\,0.91,\,0.90\\
		\hline 	
		2.5 & 1.00,\,1.00,\,1.00 & 1.07,\,1.12,\,1.14 & 1.37,\,0.70,\,0.72 \\
		\hline
		3.0 & 1.00,\,1.00,\,1.00 & 1.07,\,1.00,\,1.02 & 1.04,\,0.61,\,0.63 \\
		\hline
	\end{tabularx} 
	\caption{Relative actual cost for subsystems 1, 2 and 3.}
	\label{tab:actual cost ratio}
	\bigskip
	
	\centering
	\begin{tabularx}{0.5\textwidth}{|>{\centering\arraybackslash\hsize=0.4\hsize\linewidth=\hsize}X||>{\centering\arraybackslash\hsize=1.2\hsize\linewidth=\hsize}X|>{\centering\arraybackslash\hsize=1.2\hsize\linewidth=\hsize}X|>{\centering\arraybackslash\hsize=1.2\hsize\linewidth=\hsize}X|}
		\hline 
		$ \xi_{11} $ & Proposed DMPC & DMPC with fixed reference  & Sequential DMPC \cite{Nikou2019}  \\ 
		\hline
		1.5 & 0.0261 & 0.0255 & 0.1116\\
		\hline
		2.0 & 0.0267 & 0.0266 & 0.1155\\
		\hline 	
		2.5 & 0.0254 & 0.0253 & 0.1106 \\
		\hline 	
		3.0 & 0.0265 & 0.0262 & 0.1157 \\
		\hline
	\end{tabularx} 
	\caption{Average computational times for calculating control inputs in seconds.}
	\label{tab:computation times}
\end{table}

The simulation results for the relative actual costs and computational times are summarized in Tables~\ref{tab:actual cost ratio} and~\ref{tab:computation times}. For robot $ i $, the actual cost is computed over the simulated time interval as $ J^{a}_{i} = \sum_{\kappa = 0}^{T_{\text{sim}}} ||\hat{\mathbf{x}}_i[\kappa]-\xi_i||_{\bm{Q}_i} + ||\hat{u}_i[\kappa]-u_{\xi_{i}}||_{\bm{R}_{i}} $ where $ T_{\text{sim}}=200 $; the $ i $-th entry in each field of Table~\ref{tab:actual cost ratio} is the actual cost $ J^{a}_{i} $ normed with the actual cost $ J^{a}_{i} $ of the proposed DMPC. The presented numbers are the average from 100 simulations. 

In Fig.~\ref{fig:nominal_distances}, the nominal inter-robot distances $ \hat{d}_{ij} = ||\hat{\mathbf{x}}^{\text{pos}}_{i} - \hat{\mathbf{x}}^{\text{pos}}_{j}|| $ are depicted for various $ \xi_{11} $. For the pair $ (\hat{\mathbf{x}}^{\text{pos}}_{i}, \hat{\mathbf{x}}^{\text{pos}}_{j}) $, \eqref{eq:simplified concistency constraint dist condition} can be rewritten as 
\begin{align}
	\label{eq:simplified concistency constraint dist condition2}
	\hat{d}_{ij} - d^{\text{max}} \leq -\nu_{c_{ij}}
\end{align}
which is a condition on the nominal inter-robot distance. Intuitively, this means that if the graph of $ \hat{d}_{ij} $ exceeds the dashed line in Fig.~\ref{fig:nominal_distances}, then \eqref{eq:simplified concistency constraint dist condition2} is violated. As a consequence, the reference state at the respective time is not changed in order to prevent a potential violation of the coupled state constraints (cf. Sec.~\ref{subsec:detremination of reference states}). 

This observation can be related to the relative actual costs in Table~\ref{tab:actual cost ratio}. The closer the nominal distance gets to the dashed line or even exceeds it, the closer is the performance of the proposed DMPC to that of the DMPC with fixed reference (cf. $ \xi_{11} = 3.0 $). However, if $ \hat{d}_{ij} $ does not exceed the dashed line, the performance of the proposed DMPC scheme is significantly improved compared to that of the DMPC with fixed reference where the relative costs are up to 45\% higher (cf. $ \xi_{11} = 1.5, \, 2.0 $). Compared to sequential DMPC, the performance of the proposed DMPC scheme also tends to be better in these cases. This indicates that the parallelized evaluation of the local optimization problems can be beneficial over a sequential one despite the need of a consistency constraint. In particular, the proposed DMPC computes the control inputs more than 4 times faster (Table~\ref{tab:computation times}) than sequential DMPC. This is due to the parallel evaluation of the local optimization problems in the proposed DMPC and the reduced number of constraints. This ratio further improves in favor of the proposed DMPC if more subsystems are added. 

\begin{figure*}[t]
	\centering
	\subfigure[Initialization]{
		\def\svgwidth{0.2\textwidth}
		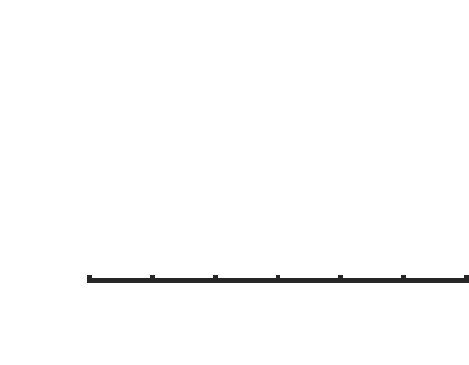
		\label{fig:iterative_init_it4}
	}\hfill
	\subfigure[Fixed reference trajectories.]{
		\def\svgwidth{0.17\textwidth}
		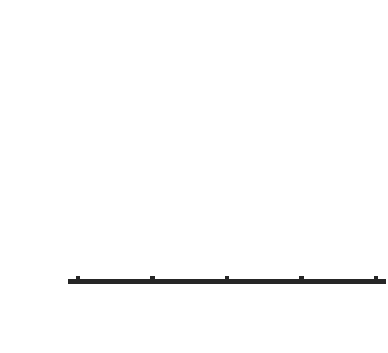
		\label{fig:iterative_traj_fixed_ref}
	}\hfill
	\subfigure[1 iteration (Algorithm~\ref{algo:DMPC})]{
		\def\svgwidth{0.17\textwidth}
		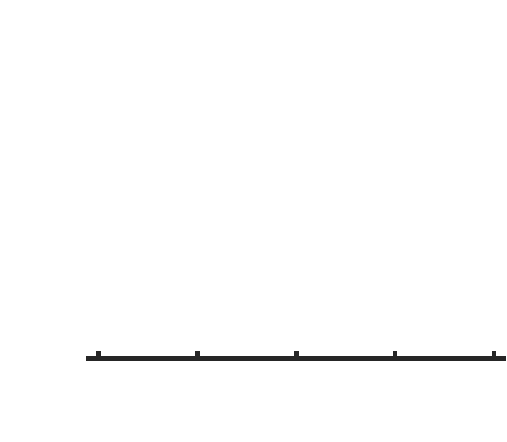
		\label{fig:iterative_traj_it1}
	}\hfill
	\subfigure[4 iterations (algorithm Rem.~\ref{remark:iterative dmpc})]{
		\def\svgwidth{0.17\textwidth}
		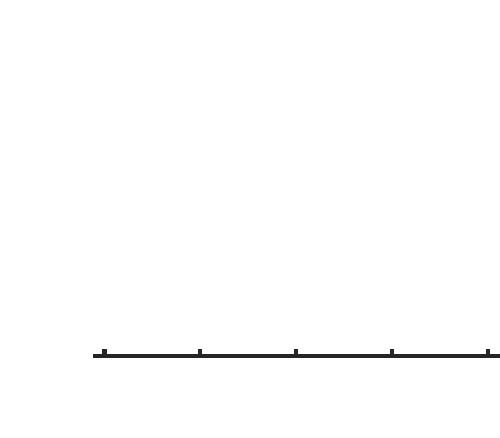
		\label{fig:iterative_traj_it4}
	}
	\vspace{-0.3cm}
	\caption{Formation control problem with collision avoidance constraints. Robots 1, 2, 3, 4 are denoted by  \textcolor{myblue}{blue}, \textcolor{mygreen}{green}, \textcolor{myred}{red}, \textcolor{myviolet}{violet}.
	}
	\label{fig:iterative}
\end{figure*}

\subsection{Collision Avoidance Constraints}

Instead of~\eqref{eq:connectivity constraint}, we now consider the collision avoidance constraint 
\begin{align}
	\label{eq:collision avoidance constraint}
	||\mathbf{x}^{\text{pos}}_{i} - \mathbf{x}^{\text{pos}}_{j}|| \geq d^{\text{min}}, \qquad j\in\calN_i:=\calV\setminus\lbrace i \rbrace
\end{align}
where $ d^{\text{min}} = 0.5 $. Therefore, we replace~\eqref{eq:simplified concistency constraint dist condition} by
\begin{align}
	\label{eq:simplified concistency constraint collision avoidance condition}
	c_{ij}({\mathbf{x}}_i,{\mathbf{x}}_j) = d^{\text{min}}-||{\mathbf{x}}^{\text{pos}}_{i} - {\mathbf{x}}^{\text{pos}}_{j}|| \leq - \nu_{c_{ij}}.
\end{align}

We consider four mobile robots governed by \eqref{eq:dynamics three wheeled robot} as before. The initial formation is $ x_{0,1} = [0,2,-\pi/2]^T $, $ x_{0,2} = [-2,0,0]^T $, $ x_{0,3} = [0,-2,\pi/2 ]^T $, $ x_{0,4} = [2,0,\pi ]^T $; the target formation is $ \xi_1 = [0,-2,-3\pi/2]^T $, $ \xi_2 = [2,0,-\pi]^T $, $ \xi_3 = [0,2,-\pi/2]^T $, $ \xi_4 = [-2,0,0]^T $. Everything else remains unchanged. 
We initialize the DMPC algorithms with reference trajectories where the robots move clockwise to their target states on the opposite side of the formation as depicted in Fig.~\ref{fig:iterative_init_it4}. Because it is generally difficult to determine optimal initially feasible reference trajectories in the presence of concave constraints, it can be beneficial to employ the iterative DMPC scheme as outlined in Rem.~\ref{remark:iterative dmpc}. As it can be seen from Table~\ref{tab:actual cost iter}, the actual cost $ J_{i}^{a} $ reduces with an increasing number of iterations. Fig.~\ref{fig:iterative} illustrates how the state trajectories of the closed-loop system improve with an increasing number of iterations. Observe that even though constraint~\eqref{eq:collision avoidance constraint} is non-convex, the local optimization problems in our proposed approach are only subject to convex state constraints. That is because the satisfaction of all state constraints is ensured by the consistency constraint which is convex by choice.

\begin{table}[tp]
	\centering
	\begin{tabularx}{0.49\textwidth}{|>{\centering\arraybackslash\hsize=1.2\hsize\linewidth=\hsize}X||>{\centering\arraybackslash\hsize=0.95\hsize\linewidth=\hsize}X|>{\centering\arraybackslash\hsize=0.95\hsize\linewidth=\hsize}X|>{\centering\arraybackslash\hsize=0.95\hsize\linewidth=\hsize}X|>{\centering\arraybackslash\hsize=0.95\hsize\linewidth=\hsize}X|}
		\hline 
		iterations & Robot 1 & Robot 2 & Robot 3 & Robot 4   \\ 
		\hline
		Fixed ref. & 5.2028 & 5.1523 & 5.1659 & 5.1530 \\
		\hline
		1 & 2.4460 & 2.4502 & 2.4832 & 2.4474 \\
		\hline
		4 & 1.7965 & 1.7948 & 1.7609 & 1.7633 \\
		\hline
		6 & 1.4752 & 1.4779 & 1.5088 & 1.5011 \\
		\hline
	\end{tabularx} 
	\caption{Actual cost $ J_{i}^{a} $ ($ \times 10^{4} $) in dependence of the number of iterations, average of 100 simulations. Obtained for $ T_{\text{sim}} = 100 $.}
	\label{tab:actual cost iter}

\end{table}


\section{Conclusion}
\label{sec:conclusion}

We presented a robust DMPC algorithm that allows for the parallel evaluation of the local optimization problems in the presence of coupled state constraints while it admits to alter and improve already established reference trajectories. For the case of dynamically decoupled systems subject to coupled constraints, we thereby provide a novel DMPC scheme that allows for a faster distributed control input computation compared to sequential DMPC schemes. Theoretical guarantees on recursive feasibility and robust asymptotic convergence are provided. Moreover, we briefly commented on an iterative extension of the algorithm. In the end, we demonstrated the algorithm's applicability and compared its performance to other DMPC algorithms.


\bibliographystyle{ieeetr}
\bibliography{/Users/wiltz/CloudStation/JabBib/Research/000_MyLibrary}


\balance

\begin{wrapfigure}{l}{20mm}
	\includegraphics[width=1in,height=1.25in,clip,keepaspectratio]{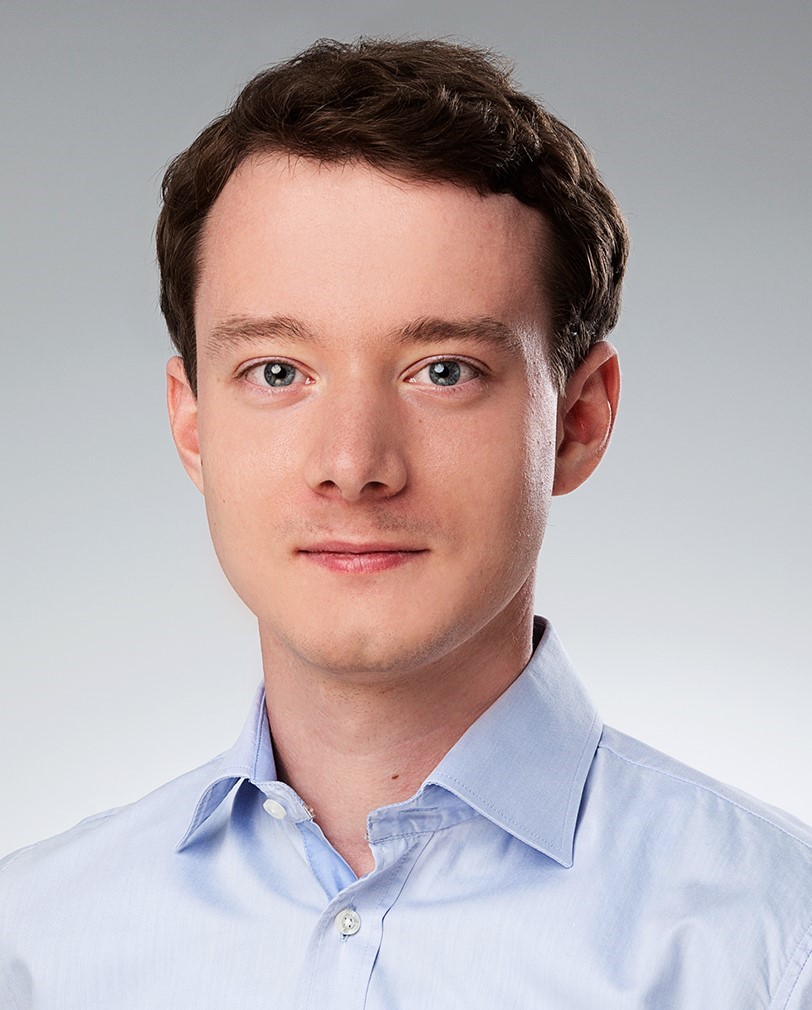}
\end{wrapfigure}\par
\textbf{Adrian Wiltz} received his B.Sc. and M.Sc. degree in Engineering Cybernetics from the University of Stuttgart (Germany) in 2017 and 2020, respectively, and is currently pursuing the Ph.D. degree at KTH Royal Institute of Technology, Stockholm (Sweden). In 2020, he worked as an intern in the field of wind turbine control at sowento GmbH, Stuttgart. His research interests are in distributed and hybrid control algorithms and the control of heterogeneous multiagent systems.

\begin{wrapfigure}{l}{20mm}
	\includegraphics[width=1in,height=1.25in,clip,keepaspectratio]{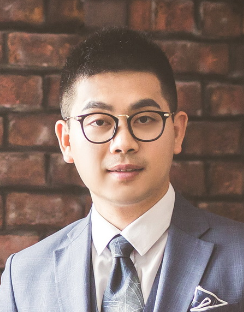}
\end{wrapfigure}\par
\textbf{Fei Chen} received his Ph.D. degree in Electrical Engineering from KTH Royal Institute of Technology, Sweden, in 2023. He received his M.Sc. degree from the Systems and Control group in the Electrical Engineering Department at Eindhoven University of Technology (Netherlands) in 2016, and his B.Sc. degree in the Department of Control Science and Engineering at Zhejiang University (China) in 2014. He is currently a postdoctoral scholar with the Department of Mechanical and Aerospace Engineering, University of California, San Diego, CA, USA. His research interests lie at the intersection of systems and control theory, multi-agent systems, optimization, and formal methods. Dr. Chen was a finalist for the Interactive Paper Prize at the IFAC World Congress 2023.

\begin{wrapfigure}{l}{20mm}
	\includegraphics[width=1in,height=1.25in,clip,keepaspectratio]{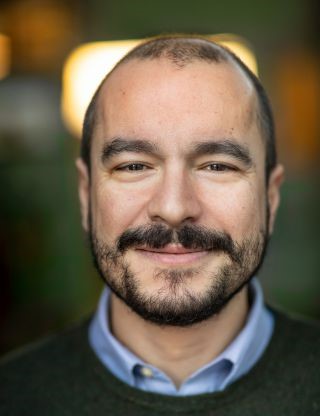}
\end{wrapfigure}\par
\textbf{Dimos V. Dimarogonas} was born in Athens, Greece, in 1978. He received the Diploma degree in electrical and computer engineering and the Ph.D. degree in mechanical engineering from the National Technical University of Athens, Athens, Greece, in 2001 and 2007, respectively. Between 2007 and 2010, he held Postdoctoral 	positions at KTH and MIT. He is currently a Professor with the Division of Decision and Control Systems, School of Electrical Engineering and Computer Science, KTH Royal Institute of Technology. His current research interests include multi-agent systems, hybrid systems and control, robot navigation and manipulation, human–robot interaction, and networked control. Prof. Dimarogonas serves as an Associate Editor of Automatica and a Senior Editor of IEEE Transactions on Control of Network Systems. He was a recipient of the ERC starting Grant in 2014, the ERC Consolidator Grant in 2019, and the Knut och Alice Wallenberg Academy Fellowship in 2015. Prof. Dimarogonas is a Fellow of the IEEE.

\end{document}